\documentclass[superscriptaddress,aps,pra,onecolumn,showkeys]{revtex4-2}
\usepackage{amsmath,mathtools,yhmath,siunitx}
\usepackage{amssymb}
\usepackage{graphicx}
\usepackage{color}
\usepackage{bm}
\usepackage{orcidlink}

\begin{document}
\title{Multifunctional Nonreciprocal Quantum Device Based on Superconducting Quantum Circuit}

\author{Yue Cai}
\affiliation{Lanzhou Center for Theoretical Physics, Key Laboratory of Theoretical Physics of Gansu Province, Key Laboratory of Quantum Theory and Applications of MoE, Gansu Provincial Research Center for Basic Disciplines of Quantum Physics, Lanzhou University, Lanzhou 730000, China}

\author{Jie Liu}
\affiliation{Lanzhou Center for Theoretical Physics, Key Laboratory of Theoretical Physics of Gansu Province, Key Laboratory of Quantum Theory and Applications of MoE, Gansu Provincial Research Center for Basic Disciplines of Quantum Physics, Lanzhou University, Lanzhou 730000, China}

\author{Kang-Jie Ma}
\affiliation{Lanzhou Center for Theoretical Physics, Key Laboratory of Theoretical Physics of Gansu Province, Key Laboratory of Quantum Theory and Applications of MoE, Gansu Provincial Research Center for Basic Disciplines of Quantum Physics, Lanzhou University, Lanzhou 730000, China}

\author{Lei Tan}
\email{tanlei@lzu.edu.cn}
\affiliation{Lanzhou Center for Theoretical Physics, Key Laboratory of Theoretical Physics of Gansu Province, Key Laboratory of Quantum Theory and Applications of MoE, Gansu Provincial Research Center for Basic Disciplines of Quantum Physics, Lanzhou University, Lanzhou 730000, China}

\begin{abstract}
Nonreciprocal devices, such as isolator or circulator, are crucial for information routing and processing in quantum networks. Traditional nonreciprocal devices, which rely on the application of bias magnetic fields to break time-reversal symmetry and Lorentz reciprocity, tend to be bulky and require strong static magnetic fields. This makes them challenging to implement in highly integrated large-scale quantum networks. Therefore, we design a multifunctional nonreciprocal quantum device based on the integration and tunable interaction of superconducting quantum circuit. This device can switch between two-port isolator, three-port symmetric circulator, and antisymmetric circulator under the control of external magnetic flux. Furthermore, both isolator and circulator can achieve nearly perfect unidirectional signal transmission. We believe that this scalable and integrable nonreciprocal device could provide new insight for the development of large-scale quantum networks.

\end{abstract}

\maketitle

\section{INTRODUCTION}
Quantum networks represent an emerging paradigm in information processing\cite{quantum_network_1,quantum_network_2,quantum_network_3,quantum_network_4,quantum_network_5}, playing an indispensable role in the implementation and interconnection of strategic frontier technology that combines quantum physics and information technology\cite{quantum_technology}, such as quantum computing\cite{quantum_computing_1}, quantum communication\cite{quantum_communication_1,quantum_communication_2,quantum_communication_3,quantum_communication_4}, quantum radar\cite{quantum_radar_1,quantum_radar_2,quantum_radar_3}, and quantum measurement\cite{quantum_measurement_1,quantum_measurement_2,quantum_measurement_3}. Nonreciprocal optical devices, such as isolator\cite{isolator_1} or circulator\cite{circulator_1}, serve as fundamental components for information routing and processing in quantum networks. These devices are crucial for protecting fragile signals from harmful backflow noise. Traditional methods for creating nonreciprocity rely on the application of magnetic bias fields to break time-reversal symmetry and Lorentz reciprocity\cite{isolator_1,symmetry_breaking_1,symmetry_breaking_2,symmetry_breaking_3}, typically using ferrite materials\cite{ferrite_1,ferrite_2,ferrite_3} and Faraday effect\cite{faraday_effect_1,faraday_effect_2}. However, these devices are often bulky and require strong static magnetic fields. This makes them difficult to implement in highly integrated large-scale quantum networks.

In recent years, remarkable progress has been made in superconducting quantum circuit (SQC) due to their potential applications in quantum information processing and microwave photonics\cite{SQC_1,SQC_2,SQC_3,SQC_4,SQC_5}. The primary advantages of these on-chip structures include scalability, integration, and tunable strong interactions between multiple modes. Those are highly suitable for the development of devices that consider integrated design. 
Therefore, in this paper, we propose a nonreciprocal transmission scheme based on SQC. It consists of two-level superconducting artificial atom (A-A) and microwave resonant cavities. In our design, A-A and cavity are replaced by transmon qubit\cite{A_A_1,A_A_2,A_A_3} and LC oscillating circuits\cite{LCC_1,LCC_2,LCC_3,LCC_4}, respectively. By using superconducting quantum interference device (SQUID)\cite{SQUID_1,SQUID_2,SQUID_3}, pierced by external magnetic flux, to connect different integrated elements, tunable coupling between different modes can be achieved. 
Our scheme enables nearly perfect unidirectional signal transmission and completely suppressing transmission in the opposite direction, i.e. an isolator, when the intrinsic damping rates of various modes are considered. Furthermore, our proposal offers both operational convenience and functional scalability. Specifically, by controlling the external magnetic flux, the multifunctional nonreciprocal quantum device can switch between two-port isolator, three-port symmetric-circulator, and antisymmetric-circulator. The direction of signal transmission in the isolator or circulator can also be conveniently determined by external magnetic flux. 
Due to space limitations, this paper focuses on only three basic nonreciprocal transmission functionalities as examples. In fact, leveraging advantage of the integration and tunable interactions of SQC, additional functionalities can be incorporated into this multifunctional nonreciprocal quantum device using similar procedure. We believe that this scalable and integrable nonreciprocal device could be more practical for large-scale quantum networks.

This paper is organized as follow: In Sec. \ref{II}, we first introduce a two-port isolator model based on SQC, and provide the corresponding Hamiltonian and scattering matrix. In Sec. \ref{III}, we analyze the scattering characteristics of this isolator, showing that it can achieve nearly perfect unidirectional signal transmission when the intrinsic damping rates of various modes are taken into account. Based on the two-port isolator model, in Sec. \ref{IV}, we obtain symmetric and antisymmetric circulator with different transmission characteristics by considering additional auxiliary or transition cavities. These three nonreciprocal devices are then integrated into a multifunctional nonreciprocal quantum device in Sec. \ref{V}, which can switch between the three functionalities through the overall adjustment of external magnetic flux. Finally, a brief summary and extended discussion are provided in the conclusion section.

\section{Basic Model}\label{II}

\begin{figure}[htb]
	
	\centering
	\includegraphics[width=0.7\linewidth]{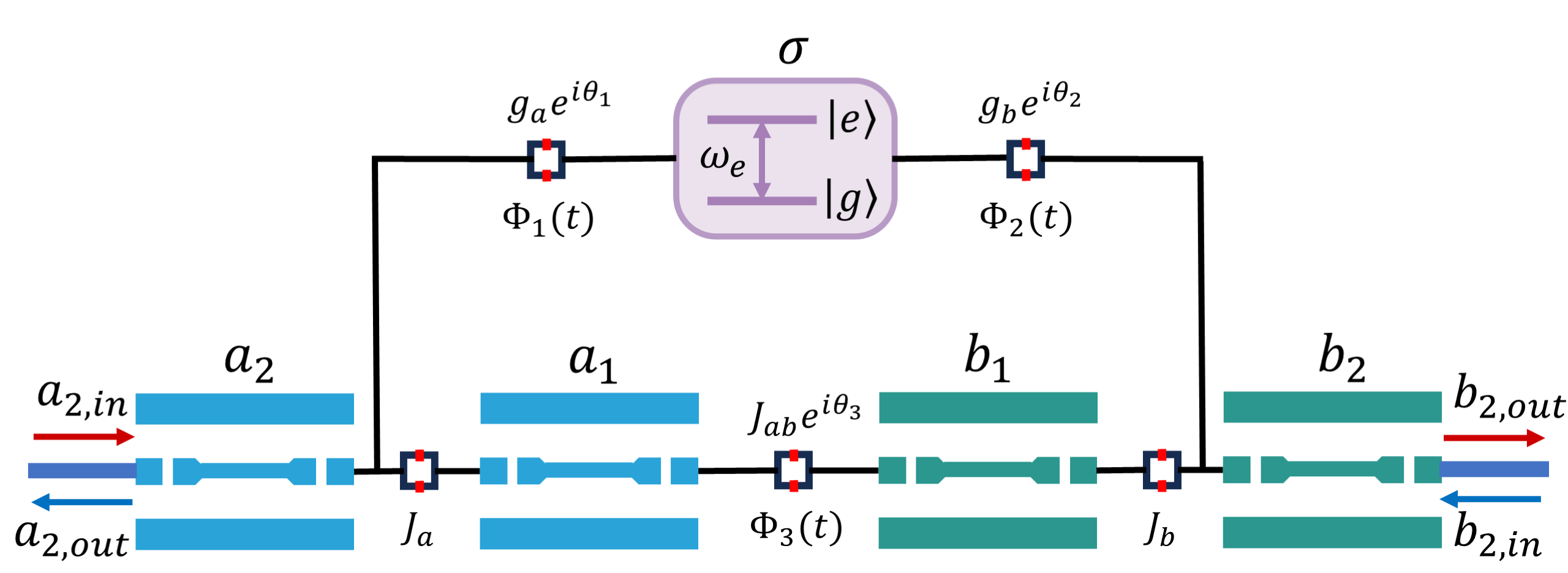}
	
	\caption{Schematic diagram of two-port isolator based on SQC, consisting of two-level superconducting A-A $\sigma$ and microwave resonant cavities $a_j$, $b_j$ ($j=1,2$). The coupling between different modes is achieved through SQUID. External magnetic flux $\Phi_1(t)$ ($\Phi_2(t)$) is applied to pierce the SQUID connecting A-A $\sigma$ with cavity mode $a_2$ ($b_2$), while $\Phi_3(t)$ is applied to the SQUID connecting cavity modes $a_1$ and $b_1$.
	}
	
	\label{fig1}
\end{figure}

As schematically shown in Fig. \ref{fig1}, the system under consideration is based on a superconducting architecture consisting of a two-level superconducting A-A $\sigma$ and four microwave resonant cavities $a_{j}$,$b_{j}$ ($j=1,2$). Specifically, the two-level A-A and cavities are experimentally replaced by transmon qubit\cite{A_A_1,A_A_2,A_A_3} and LC oscillating circuits\cite{LCC_1,LCC_2,LCC_3,LCC_4}, respectively, while SQUID is used to connect different lumped-elements, enabling coupling between different modes\cite{SQUID_1,SQUID_2,SQUID_3}. The Hamiltonian governing the dynamics of this system is given by (setting $\hbar=1$)
\begin{equation}\label{eq1}
	\begin{split}
		\hat{H}_1  = 
		\sum_{d=a,b} \sum_{j=1,2} \omega_{d} d_{j}^{\dagger}d_{j} + 
		\omega_{e} \sigma^{\dagger}\sigma + 
		(J_{a} a_{1}^{\dagger}a_{2} + J_{b} b_{1}^{\dagger}b_{2}
		+ g_a(t) a_{2}^{\dagger}\sigma + g_b(t) b_{2}^{\dagger}\sigma
		+ J_{ab}(t) a_{1}^{\dagger}b_{1} + \text{\text{H.c.}}).
	\end{split}
\end{equation}
Where $\omega_{d} (d=a,b)$ represent the intrinsic frequencies of the cavity modes $d_{j}$, while $\omega_{e}$ is the transition frequency between the ground state $|g\rangle$ and the excited state $|e\rangle$ of the A-A. $J_{a}$ ($J_{b}$) is the photon hopping rate between cavity $a_1$ and $a_2$ (or $b_1$ and $b_2$). $g_{a}(t)=2g_{a}cos(\omega_{1}t+\theta_{1})$ ($g_{b}(t)=2g_{b}cos(\omega_{2}t+\theta_{2})$) describes the time-dependent coupling between the A-A and cavities $a_2$ ($b_2$), while $J_{ab}(t)=2J_{ab}cos(\omega_{3}t+\theta_{3})$ represents the coupling between cavity modes $a_1$ and $b_1$, where $\omega_n$ and $\theta_n$ denote the modulated frequency and phase, respectively.
In the experiment, such time-dependent interaction can be implemented via the SQUID with a tunable inductance. Specifically, the SQUID consists of a superconducting loop containing two identical Josephson junctions, whose inductance variation can change the electrical boundary conditions of other modes and their interaction\cite{SQUID_4,SQUID_5}. Therefore, by applying a rapidly oscillating magnetic flux $\Delta \Phi_n \cos{(\omega_n+\theta_n)}$ piercing the loops of the SQUID, one can achieve a time-varying interaction\cite{SQUID_1,SQUID_6}.

When the frequency $\omega_n$ is chosen to match $\omega_{1}=\omega_{e}-\omega_{a}$, $\omega_{2}=\omega_{e}-\omega_{b}$, $\omega_{3}=\omega_{b}-\omega_{a}$, the rotating wave approximation can be applied to neglect rapidly oscillating terms. In the interaction picture ($\hat{H}_{1,0}=\sum_{d=a,b} \sum_{j=1,2} \omega_{d} d_{j}^{\dagger}d_{j} + \omega_{e} \sigma^{\dagger}\sigma$), the Hamiltonian of Eq. \ref{eq1} is reduced to
\begin{equation}\label{eq2}
	\hat{H}_{1,int}=J_{a} a_{1}^{\dagger}a_{2}  + J_{b} b_{1}^{\dagger}b_{2} + g_{a} a_{2}^{\dagger}\sigma + g_{b} b_{2}^{\dagger}\sigma + J_{ab} e^{i\phi_1} a_{1}^{\dagger}b_{1} + \text{\text{H.c.}}.
\end{equation}
Here, all phases are absorbed into the total phase $\phi_1\equiv\theta_{2}+\theta_{3}-\theta_{1}$, as only the total phase has physical significance\cite{nonreciprocal_graph,phase_included}. Without loss of generality, $\phi_1$ is retained only in the terms of $a_{1}^{\dagger}b_{1}$ and $b_{1}^{\dagger}a_{1}$ in Eq. \ref{eq2}. It is important to emphasize that the presence of the nontrivial phase $\phi_1$ plays a crucial role in breaking the time-reversal symmetry of the system\cite{LCC_3}, which is modulated by the external magnetic flux piercing three SQUIDs. For an appropriate phase $\phi_1$, a isolator can be created, enabling unidirectional propagation of photon as information carrier, thus protecting fragile signal from harmful backflow noise.

Using the Heisenberg equation of motion, one can derive the quantum Langevin equations (QLEs) of this system from the Hamiltonian Eq. \ref{eq2}
\begin{equation}\label{eq3}
	\begin{split}
		\dot{a}_{1} & = -i \left( J_{a} a_{2} + J_{ab} e^{i\phi_1} b_{1} \right) - \frac{1}{2} \gamma_{c} a_{1} + f_{a1}\\
		\dot{a}_{2} & = -i \left( J_{a} a_{1} + g_{a} \sigma \right) - \frac{1}{2} \left( \kappa_{c,1} + \gamma_{c} \right) a_{2} + \sqrt{\kappa_{c,1}} a_{2,in} + f_{a2}\\
		\dot{b}_{1} & = -i \left( J_{b} b_{2} + J_{ab} e^{-i\phi_1} a_{1} \right) - \frac{1}{2} \gamma_{c} b_{1} + f_{b1}\\
		\dot{b}_{2} & = -i \left( J_{b} b_{1} + g_{b} \sigma \right) - \frac{1}{2} \left( \kappa_{c,2} + \gamma_{c} \right) b_{2} + \sqrt{\kappa_{c,2}} b_{2,in} + f_{b2}\\
		\dot{\sigma} & = -i \left( g_{a} a_{2} + g_{b} b_{2} \right) - \frac{1}{2} \gamma_{e} \sigma + f_{\sigma}, 
	\end{split}
\end{equation}
where $\gamma_c$ and $\gamma_e$ are the intrinsic damping rates of cavities and A-A, respectively. $\kappa_{c,1}$ ($\kappa_{c,2}$) is the external damping rate of cavity $a_2$ ($b_2$), arising from the coupling between the cavity modes and external signal transmission channels (often modeled by transmission lines in SQC)\cite{TL_1,nonreciprocal_router}. Additionally, $a_{2,in}$ ($b_{2,in}$) represents the quantum signal input from the $a$ ($b$) port, while $f_{aj}$, $f_{bj}$, and $f_{\sigma}$ are the quantum noise operators corresponding to the respective modes.

To further solve this equation, we introduce a Fourier transform $f\left(t\right)=\int_{-\infty}^{+\infty}\frac{d\omega}{2\pi}f\left(\omega\right)e^{-i\omega t}$ to convert Eq. \ref{eq3} from the time domain to the frequency domain
\begin{equation}\label{eq4}
	\begin{split}
		\left( \frac{1}{2} \gamma_{c} - i \omega \right) a_{1}(\omega) + iJ_{a}a_{2}(\omega) + iJ_{ab}e^{i\phi_1}b_{1}(\omega) &= f_{a1}(\omega) \\
		\left[ \frac{1}{2} \left( \kappa_{c,1} + \gamma_{c} \right) - i \omega \right] a_{2}(\omega) + iJ_{a}a_{1}(\omega) + ig_{a}\sigma(\omega)  &= \sqrt{\kappa_{c,1}} a_{2,in}(\omega) + f_{a2}(\omega)\\
		\left( \frac{1}{2} \gamma_{c} - i \omega \right) b_{1}(\omega) + iJ_{b}b_{2}(\omega) + iJ_{ab}e^{-i\phi_1}a_{1}(\omega) &= f_{b1}(\omega)\\
		\left[ \frac{1}{2} \left( \kappa_{c,2} + \gamma_{c} \right) - i \omega \right] b_{2}(\omega) + iJ_{b}b_{1}(\omega) + ig_{b}\sigma(\omega)  &= \sqrt{\kappa_{c,2}} b_{2,in}(\omega) + f_{b2}(\omega)\\
		\left( \frac{1}{2} \gamma_{e} - i \omega \right) \sigma(\omega) + ig_{a}a_{2}(\omega) + ig_{b}b_{2}(\omega) &= f_{\sigma}(\omega). 
	\end{split}
\end{equation}
According to the standard input-output boundary conditions $a_{2,out}(\omega) = a_{2,in}(\omega) - \sqrt{\kappa_{c,1}} a_{2}(\omega) $, $b_{2,out}(\omega) = b_{2,in}(\omega) - \sqrt{\kappa_{c,2}} b_{2}(\omega)$, the output field can be expressed in matrix form
\begin{equation}\label{eq5}
	U_{1,out}(\omega) = S_1(\omega) U_{1,in}(\omega) + F(\omega),
\end{equation} 
where $U_{1,in}(\omega)=\left[a_{2,in}(\omega),b_{2,in}(\omega)\right]^{\mathrm{T}}$ and $U_{1,out}(\omega)=\left[a_{2,out}(\omega),b_{2,out}(\omega)\right]^{\mathrm{T}}$ represent the vectors of input and output quantum fields, respectively. $F(\omega)$ represents the vector of quantum noises. It is note that the Hamiltonian $\hat{H}_1$ only includes the beam splitter-type couplings, meaning that the output quantum fields are solely affected by thermal noises. At a low operating temperature, the average thermal photon number of each mode is very small, so their impact on the output fields can be neglected\cite{LCC_3}. For instance, when the system operates on a superconducting circuit platform at a temperature of T=20mK, the thermal excitation of the microwave cavity is $\sim 10^{-7}$, while the resonance frequency is 2$\pi \times 6$GHz. Furthermore, $S_1(\omega)$ represents the scattering matrix
\begin{equation}\label{eq6}
	S_1(\omega) = 
	\begin{bmatrix}
		r_1^{aa} & t_1^{ba}\\
		t_1^{ab} & r_1^{bb},
	\end{bmatrix}
\end{equation} 
the corresponding matrix elements are given by
\begin{equation}\label{eq7}
	\begin{split}
		t_1^{ab} &= \frac{i\sqrt{\kappa_{c,1}\kappa_{c,2}}M_{-}}{M_{+}M_{-}-F_{a}F_{b}}\\
		r_1^{aa} &= 1+\frac{i\kappa_{c,1}F_{b}}{M_{+}M_{-}-F_{a}F_{b}}\\
		t_1^{ba} &= \frac{i\sqrt{\kappa_{c,1}\kappa_{c,2}}M_{+}}{M_{+}M_{-}-F_{a}F_{b}}\\
		r_1^{bb} &= 1+\frac{i\kappa_{c,2}F_{a}}{M_{+}M_{-}-F_{a}F_{b}}, 
	\end{split}
\end{equation}
where $M_{\pm}=J_{a}J_{b}J_{ab}e^{\pm i\phi_1}/D+g_{a}g_{b}/\omega_{eff,2}$, $F_{\alpha (\alpha=a,b)}=\omega_{eff,\alpha}-\omega_{eff,1}J_{\alpha}^{2}/D-g_{\alpha}^{2}/\omega_{eff,2}$, $D=\omega_{eff,1}^{2}-J_{ab}^{2}$. $\omega_{eff,a}=\omega+i\frac{\kappa_{c,1}+\gamma_{c}}{2}$ ($\omega_{eff,b}=\omega+i\frac{\kappa_{c,2}+\gamma_{c}}{2}$), $\omega_{eff,1}=\omega+i\frac{\gamma_{c}}{2}$ and $\omega_{eff,2}=\omega+i\frac{\gamma_{e}}{2}$ are the effective frequencies of cavity modes $a_{2}$ ($b_{2}$), $a_{1}$ or $b_{1}$, and A-A when damping rate is considered, respectively. The matrix elements in Eq. \ref{eq7} describe the scattering of microwave photon through the superconducting circuit system, where $t_1^{ij}$ ($r_1^{ii}$) denotes the transmission (reflection) amplitude of the photon from input port $i$ ($i$) to output port $j$ ($i$). The nonreciprocity of this system can be demonstrated through the ratio $I_1^{ab}$ of the transmission coefficients in two opposite directions.
\begin{equation}\label{eq8}
	I_1^{ab}=\frac{t_1^{ab}}{t_1^{ba}} =
	\frac{(2\omega + i\gamma_e) J_{a}J_{b}J_{ab}e^{-i\phi_1} + 2Dg_{a}g_{b}}
	{(2\omega + i\gamma_e) J_{a}J_{b}J_{ab}e^{i\phi_1} + 2Dg_{a}g_{b}}. 
\end{equation} 
When $I_1^{ab} \neq 1$, it indicates the presence of a nonreciprocal response. From Eq. \ref{eq8}, it is evident that this nonreciprocal phenomenon arises from the combined effect of the nontrivial phase $\phi_1$ and the intrinsic damping rate $\gamma_e$ of A-A, both of which are essential. The vanishing of the phase $\phi_1$ would result in $t_1^{ab}=t_1^{ba}$, while the total damping rate of 0 for mode $\sigma$ lead to $t_1^{ab}=(t_1^{ba})^{\ast}$. Any of these two cases would cause the loss of the nonreciprocal response. Therefore, both phase and damping rate play crucial roles in violating reciprocity, as they break the symmetry between scattering elements which are linked by complex conjugation\cite{nonreciprocal_graph}. Physically, this is related to the time-reversal symmetry breaking of the system\cite{TRSB_1,TRSB_2}. When $\ensuremath{\left|t_1^{ab}\right|}=1,\ensuremath{\left|t_1^{ba}\right|}=0$ or $\ensuremath{\left|t_1^{ab}\right|}=0,\ensuremath{\left|t_1^{ba}\right|}=1$, the optimal nonreciprocity can be achieved.

\section{Two-Port Isolator}\label{III}

We now consider the conditions for achieving optimal nonreciprocity in the isolator model shown in Fig. \ref{fig1}. We first consider the implementation of unidirectional transmission from port $a$ to $b$, which requires that $M_{+}=0$ according to Eq. \ref{eq7}, i.e.,
\begin{equation}\label{eq9}
	e^{i\phi_1}=\frac{g_{a}g_{b}} {\omega_{eff,2}} \frac{J_{ab}^{2}-\omega_{eff,1}^{2}} {J_{a}J_{b}J_{ab}} 
\end{equation} 
For simplicity, we assume that A-A is symmetrically coupled to both port cavities ($g_{a}=g_{b}\equiv g$) and the intrinsic damping rate of the cavities is much smaller than the coupling strengths, such that it can be neglected (i.e. $\gamma_{c}=0$). Under these conditions, we find that transmission from port $b$ to $a$ vanishes at $\omega=0$ if
\begin{equation}\label{eq10}
	\begin{split}
		g^{2} &=\frac{J_{a}J_{b}\gamma_{e}}{2J_{ab}}\\
		 \phi &=\frac{3}{2}\pi.
	\end{split}  
\end{equation}   
Substituting condition Eq. \ref{eq10} into Eq. \ref{eq7}, the scattering matrix element $t_1^{ab}$ for unidirectional transmission is simplified to 
\begin{equation}\label{eq11}
	t_1^{ab}=\frac{8J_{a}J_{b}J_{ab}\sqrt{\kappa_{c,1} \kappa_{c,2}}}
	{\left( 2J_{a}J_{b} + J_{ab}\kappa_{c,1} \right) 
	 \left( 2J_{a}J_{b} + J_{ab}\kappa_{c,2} \right)}. 
\end{equation}  
Thus this isolator can achieve optimal unidirectional transmission from $a$ to $b$ (i.e., $\left|t_1^{ab}\right|=1, \left|t_1^{ba}\right|=0$) when the parameters are adjusted to satisfy
\begin{equation}\label{eq12}
	J_{ab}=\frac{2J_{a}J_{b}}{\kappa_c}.
\end{equation} 
For convenience, we set $\kappa_{c,1} = \kappa_{c,2} \equiv \kappa_c$ in the derivation and the condition Eq. \ref{eq10} simplifies to 
\begin{equation}\label{eq13}
	g = \frac{\sqrt{\gamma_{e}\kappa_c}}{2}.
\end{equation}

\begin{figure}[htb]
	
	\includegraphics[width=12cm]{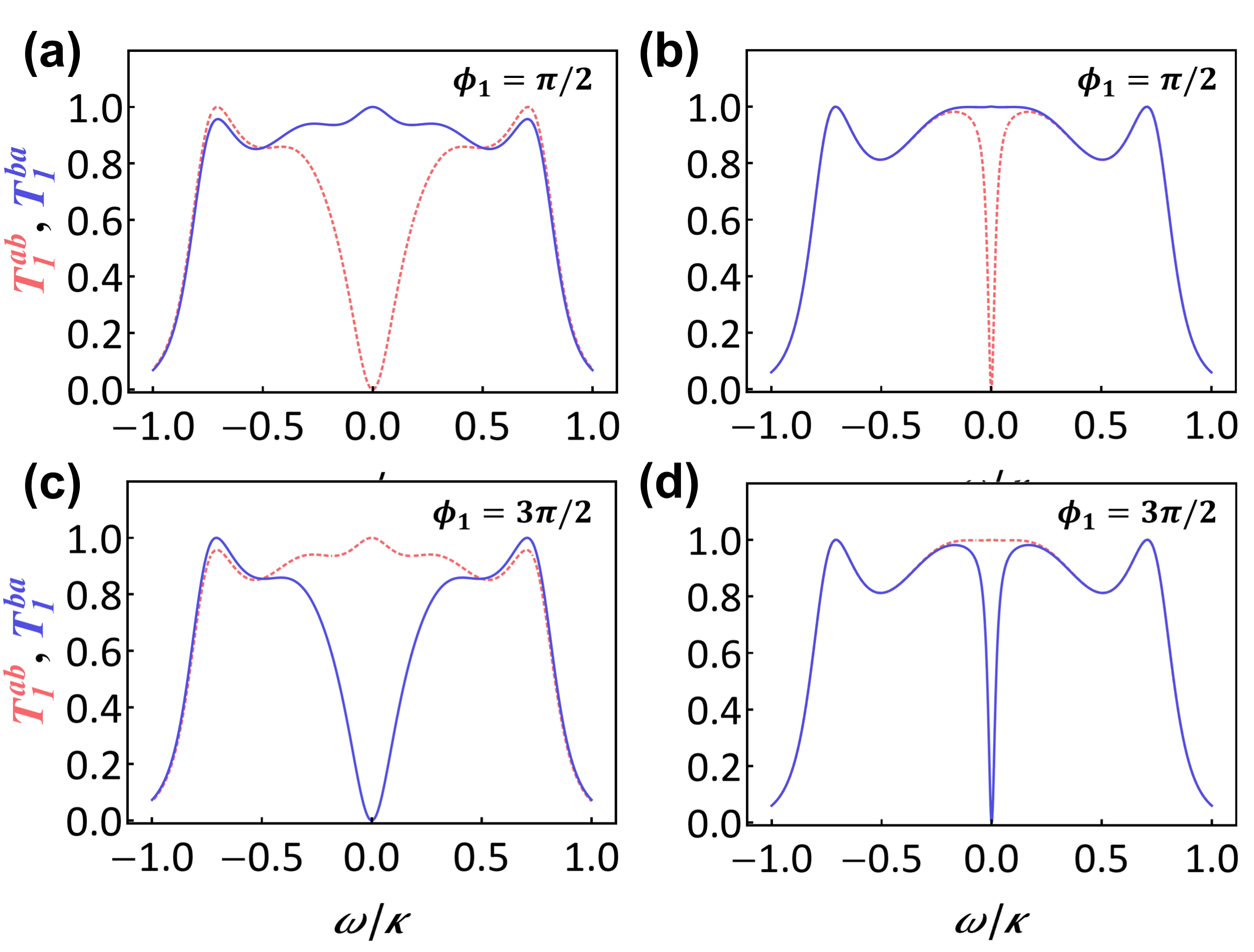}
	\centering
	
	\caption{Transmission probability $T_1^{ab}$ and $T_1^{ba}$ as a functions of frequency $\omega/\kappa$ for the different phase. 
		(a) $\gamma_e=0.15\kappa$, $\theta=\pi /2 $;
		(b) $\gamma_e=0.02\kappa$, $\theta=\pi /2 $;
		(c) $\gamma_e=0.15\kappa$, $\theta=3\pi /2 $;
		(d) $\gamma_e=0.02\kappa$, $\theta=3\pi /2 $;
		The other parameters are set to $\kappa_c=\kappa$, $J_a=J_b=\kappa/2$, while $J_{ab}$ and $g$ are given by Eqs. \ref{eq12} and \ref{eq13}.}
	
	\label{fig2}
\end{figure}

From Eq. \ref{eq7}, the transmission probability $T_1^{ab}=\left|t_1^{ab}\right|^{2}$ for photon traveling from input port $a$ to output port $b$, and the transmission probability $T_1^{ba}=\left|t_1^{ba}\right|^{2}$ for the reverse direction, are shown as a function of the frequency $\omega/\kappa$ under different phase $\phi_1$ and intrinsic damping rate $\gamma_e$ in Fig. \ref{fig2}. Where the coupling $J_{ab}$ between cavity modes $a_1$ and $b_1$, and the coupling $g$ between A-A and the port cavities are set to satisfy the conditions given by Eqs. \ref{eq12} and \ref{eq13}, respectively.
It can be observed that the presence of phase $\phi_{1}$ ($\phi_{1}\neq n\pi$) and intrinsic damping rate $\gamma_{e}$ break the time-reversal symmetry of the system, leading to the emergence of nonreciprocal phenomena\cite{TRSB_1,TRSB_2}. When the phase is $\pi/2$ or $3\pi/2$, optimal nonreciprocal response can be achieved at $\omega=0$. Specifically, as shown in Figs. \ref{fig2}(a) and \ref{fig2}(b), when the phase is $\pi/2$, this system achieves optimal unidirectional transmission from port $b$ to $a$ at $\omega=0$ (i.e., $T_1^{ba}=1$ and $T_1^{ab}=0$). Conversely, when the phase is adjusted to $3\pi/2$, unidirectional transmission in the opposite direction is implemented (i.e., $T_1^{ab}=1$ and $T_1^{ba}=0$), as depicted in Figs. \ref{fig2}(c) and \ref{fig2}(d). This indicates that by tuning the external magnetic flux, which determines the phase difference, one can conveniently select the transmission direction of this isolator. Comparing Figs. \ref{fig2}(a) and \ref{fig2}(b) (or \ref{fig2}(c) and \ref{fig2}(d)), it is evident that the bandwidth of nonreciprocity narrows as $\gamma_{e}$ decreases, but the system always exhibits optimal nonreciprocity at $\omega=0$, consistent with the theoretical analysis presented earlier.

\begin{figure}[htb]
	
	\centering
	\includegraphics[width=0.7\linewidth]{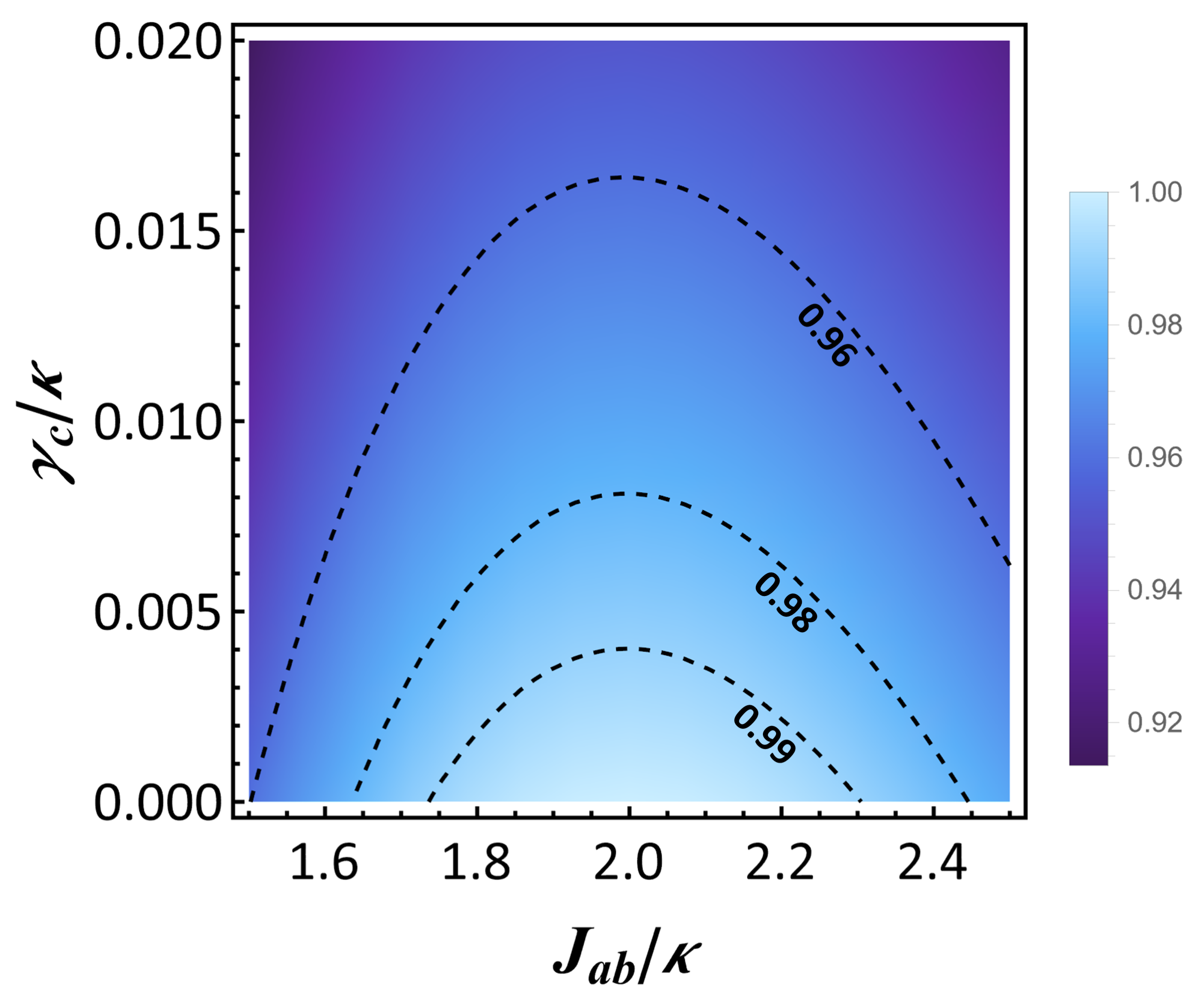}
	
	\caption{Transmission probability $T_1^{ab}$ varies with the intrinsic damping rate $\gamma_c$ and coupling $J_{ab}$ of the cavity, while the $T_1^{ba}$ in the opposite direction is 0. The parameters are set to $\kappa_c=\kappa$, $J_a=J_b=\kappa$, $\gamma_e=0.1\kappa$, $\omega=0$, and $g$ is given by Eq. \ref{eq14}.}
	
	\label{fig3}
\end{figure}

In the practical situation, the intrinsic photonic damping rate of the microwave resonators are inevitable to affect the routing performance. 
To achieve zero transmission probability in the $ b \longrightarrow a $ direction at $\omega=0$, the condition in Eq. \ref{eq10} needs to be modified to
\begin{equation}\label{eq14}
	g^{2} = \frac{2J_{a}J_{b}J_{ab}\gamma_{e}}{4J_{ab}^{2}+\gamma_{c}^{2}},
\end{equation}
while the phase $\phi_1$ remains unchanged. Substituting the condition from Eq. \ref{eq14} and the phase $\phi_1=3\pi/2$ into Eq. \ref{eq7}, the scattering matrix element $t_1^{ab}$ considering the intrinsic damping rate of cavities is derived as
\begin{equation}\label{eq15}
	\begin{split}
		t_1^{ab}
		=\frac{4J_{0}J_{1}\kappa_c}
		{\left( J_1 + 4\gamma_cJ_a^2 + J_0\kappa_t \right) 
			\left( J_1 + 4\gamma_cJ_b^2 + J_0\kappa_t \right)}.
	\end{split}
\end{equation}
Here, we similarly set $\kappa_{c,1} = \kappa_{c,2} \equiv \kappa_c$, and define $J_{0}=4J_{ab}^{2}+\gamma_{c}^{2}$, $J_{1}=8J_{a}J_{b}J_{ab}$, while $\kappa_t = \kappa_c + \gamma_c$ is the total damping rate rate of the port cavities. Therefore, the transmission probability $T_1^{ab}$ as a function of the intrinsic damping rate $\gamma_e$ and the coupling $J_{ab}$ is plotted in Fig. \ref{fig3} according to Eq. \ref{eq15}. Where the coupling $g$ between A-A and the cavities satisfies Eq. \ref{eq14}, ensuring that the reverse transmission probability $T_1^{ba} = 0$. 
As shown in Fig. \ref{fig3}, the transmission probability $T_1^{ab}$ gradually decreases with increasing intrinsic damping rate $\gamma_c$. However, by adjusting the coupling $J_{ab}$ appropriately, a high-performance isolator can still be achieved. For $\gamma_c < 0.004\kappa$, a transmission probability $T_1^{ab}>0.99$ can be realized over a large parameter range, meaning that an almost perfect single-photon isolator can be achieved. If we set $\kappa/2\pi=20$MHz, the intrinsic damping rate only needs to satisfy $\gamma_c<80$KHz to achieve $T_1^{ab} > 0.99$. This requires an internal quality factor $Q_i=\omega_i/\gamma_c > 0.75 \times 10^5$ for a cavity with resonance frequency $\omega_i = 6$GHz. Experimentally, superconducting cavities with an internal quality factor $Q_i > 10^7$ have been reported\cite{Hcavity_1,Hcavity_2}. Therefore, this isolator scheme is applicable to current available technologies, and the intrinsic damping rate $\gamma_c$ of the cavity will not be considered further in the following.

\section{Three-Port Symmetric or Antisymmetric Circulator}\label{IV}

\begin{figure}[htb]
	
	\centering
	\includegraphics[width=0.7\linewidth]{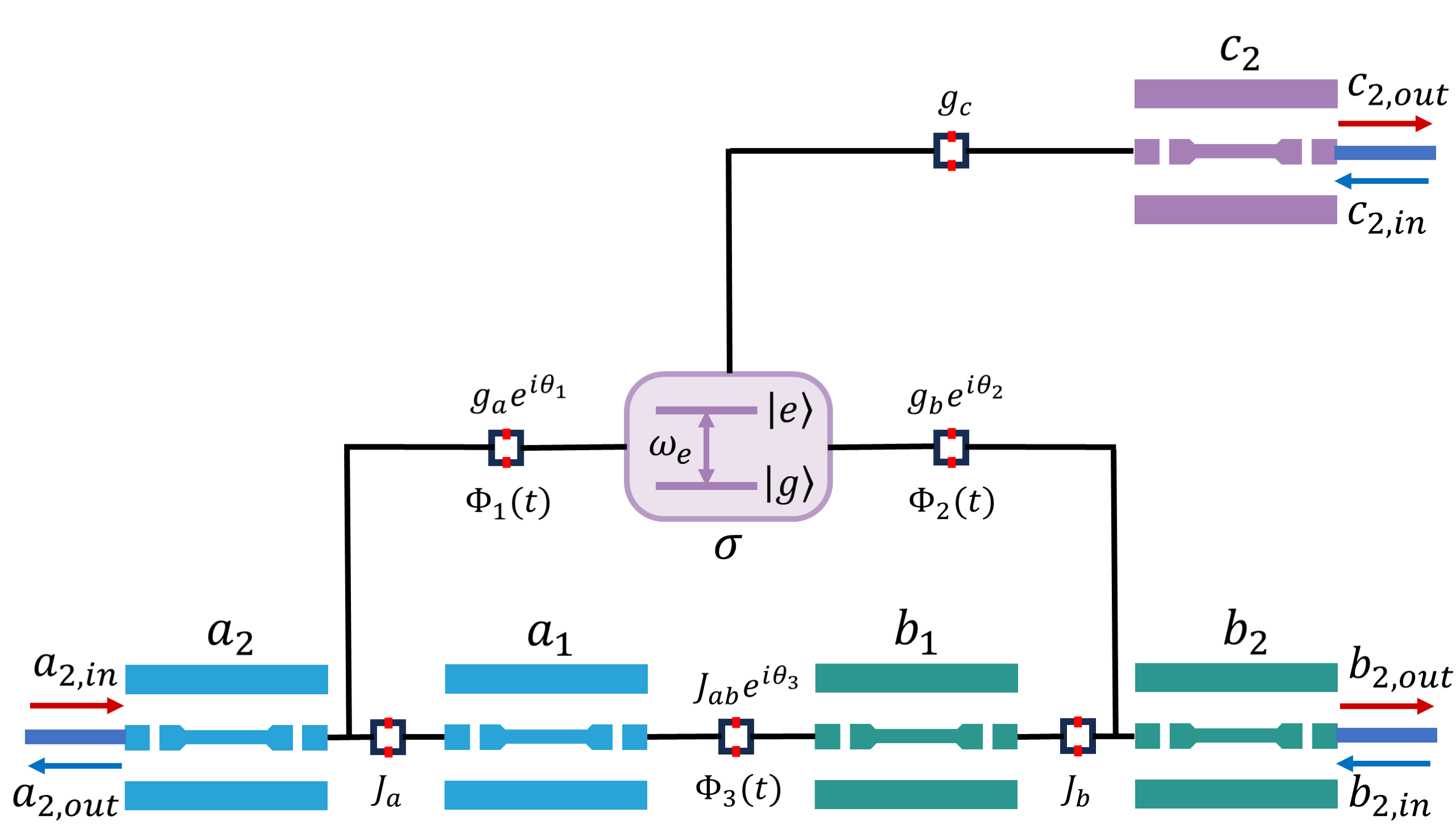}
	
	\caption{Schematic diagram of a three-port ($a_2$, $b_2$, and $c_2$) symmetric-circulator composed of a two-level superconducting A-A and superconducting resonant cavities. An auxiliary cavity $c_2$ that is coupled only to A-A is additionally added to the system.
	}
	
	\label{fig4}
\end{figure}

In this section, we design a three-port symmetric or antisymmetric circulator based on the nonreciprocity discussed in Sec. \ref{III}. To provide the third port connecting to the external signal, an auxiliary cavity $c_2$, which couples only to the A-A $\sigma$, is incorporated into our system, as shown in Fig. \ref{fig4}. The Hamiltonian of this modified system is given by
\begin{equation}\label{eq16}
	\hat{H}_2 = \hat{H}_1 + \hat{H}_{aux},
\end{equation}
and
\begin{equation}\label{eq17}
	\hat{H}_{aux} = \omega_e c_2^{\dagger} c_2 + g_c \left( c_2^{\dagger} \sigma + c_2 \sigma^{\dagger} \right)
\end{equation} 
where $c_2$ ($c_2^{\dagger}$) represents the bosonic annihilation (creation) operator for the auxiliary cavity mode with a resonance frequency $\omega_e$, and $g_c$ is the tunable coupling between the cavity mode $c_2$ and the A-A. In the interaction picture respect to $\hat{H}_{2,0}= \hat{H}_{1,0} + \omega_{e} c_2^{\dagger}c_2$, the Hamiltonian in Eq. \ref{eq16} can be written as
\begin{equation}\label{eq18}
	\hat{H}_{2,int} = \hat{H}_{1,int} + g_c \left(  c_2^{\dagger} \sigma + c_2 \sigma^{\dagger} \right).
\end{equation}
Then the QLEs for this modified system are given by
\begin{equation}\label{eq19}
	\begin{split}
		\dot{a}_{1} & = -i \left( J_{a} a_{2} + J_{ab} e^{i\phi_1} b_{1} \right) - \frac{1}{2} \gamma_{c} a_{1}\\
		\dot{a}_{2} & = -i \left( J_{a} a_{1} + g_{a} \sigma \right) - \frac{1}{2} \left( \kappa_{c,1} + \gamma_{c} \right) a_{2} + \sqrt{\kappa_{c,1}} a_{2,in}\\
		\dot{b}_{1} & = -i \left( J_{b} b_{2} + J_{ab} e^{-i\phi_1} a_{1} \right) - \frac{1}{2} \gamma_{c} b_{1}\\
		\dot{b}_{2} & = -i \left( J_{b} b_{1} + g_{b} \sigma \right) - \frac{1}{2} \left( \kappa_{c,2} + \gamma_{c} \right) b_{2} + \sqrt{\kappa_{c,2}} b_{2,in}\\
		\dot{c}_{2} & = -i g_{c} \sigma  - \frac{1}{2} \left( \kappa_{c,3} + \gamma_{c} \right) c_{2} + \sqrt{\kappa_{c,3}} c_{2,in}\\
		\dot{\sigma} & = -i \left( g_{a} a_{2} + g_{b} b_{2} + g_{c} c_{2} \right) - \frac{1}{2} \gamma_{e} \sigma. 
	\end{split}
\end{equation}  
By using Fourier transforms and standard input-output boundary conditions $a_{2,out}(\omega) = a_{2,in}(\omega) - \sqrt{\kappa_{c,1}} a_{2}(\omega) $, $b_{2,out}(\omega) = b_{2,in}(\omega) - \sqrt{\kappa_{c,2}} b_{2}(\omega) $, $c_{2,out}(\omega) = c_{2,in}(\omega) - \sqrt{\kappa_{c,3}} c_{2}(\omega) $, the output field can be expressed in matrix form
\begin{equation}\label{eq20}
	U_{2,out}(\omega) = S_{2}(\omega) U_{2,in}(\omega).
\end{equation} 
Here, $U_{2,in}(\omega)=\left[a_{2,in}(\omega),b_{2,in}(\omega),c_{2,in}(\omega)\right]^{\mathrm{T}}$ and $U_{2,out}(\omega)=\left[a_{2,out}(\omega),b_{2,out}(\omega),c_{2,out}(\omega)\right]^{\mathrm{T}}$ represent the vectors of input and output quantum fields, respectively. $S_{2}(\omega)$ is the scattering matrix
\begin{equation}\label{eq21}
	S_{2}(\omega) = 
	\begin{bmatrix}
		r_2^{aa} & t_2^{ba} & t_2^{ca}\\
		t_2^{ab} & r_2^{bb} & t_2^{cb}\\
		t_2^{ac} & t_2^{bc} & r_2^{cc}
	\end{bmatrix},
\end{equation}
where the transmission matrix elements are given by
\begin{equation}\label{eq22}
	\begin{split}
		t_2^{ab} &= \frac
		{i (F_cM_{-} + G_{ac}G_{bc}) \sqrt{\kappa_{c,1}\kappa_{c,2}}}
		{F_c (M_{+}M_{-}-F_{a}F_{b}) + G_{bc}^2 F_a + G_{ac}^2 F_b + G_{bc}G_{bc}(M_{+}+M_{-})}\\
		t_2^{ba} &= \frac
		{i (F_cM_{+} + G_{ac}G_{bc}) \sqrt{\kappa_{c,1}\kappa_{c,2}}}
		{F_c (M_{+}M_{-}-F_{a}F_{b}) + G_{bc}^2 F_a + G_{ac}^2 F_b + G_{bc}G_{bc}(M_{+}+M_{-})}\\
		t_2^{ac} &= \frac
		{i (F_bG_{ac} + M_{-}G_{bc}) \sqrt{\kappa_{c,1}\kappa_{c,3}}}
		{F_c (M_{+}M_{-}-F_{a}F_{b}) + G_{bc}^2 F_a + G_{ac}^2 F_b + G_{bc}G_{bc}(M_{+}+M_{-})}\\
		t_2^{ca} &= \frac
		{i (F_bG_{ac} + M_{+}G_{bc}) \sqrt{\kappa_{c,1}\kappa_{c,3}}}
		{F_c (M_{+}M_{-}-F_{a}F_{b}) + G_{bc}^2 F_a + G_{ac}^2 F_b + G_{bc}G_{bc}(M_{+}+M_{-})}\\
		t_2^{bc} &= \frac
		{i (F_aG_{bc} + M_{+}G_{ac}) \sqrt{\kappa_{c,2}\kappa_{c,3}}}
		{F_c (M_{+}M_{-}-F_{a}F_{b}) + G_{bc}^2 F_a + G_{ac}^2 F_b + G_{bc}G_{bc}(M_{+}+M_{-})}\\
		t_2^{cb} &= \frac
		{i (F_aG_{bc} + M_{-}G_{ac}) \sqrt{\kappa_{c,2}\kappa_{c,3}}}
		{F_c (M_{+}M_{-}-F_{a}F_{b}) + G_{bc}^2 F_a + G_{ac}^2 F_b + G_{bc}G_{bc}(M_{+}+M_{-})}\\
		r_2^{aa} &= 1 + \frac
		{i (F_bF_{c} - G_{bc}^2) \kappa_{c,1}}
		{F_c (M_{+}M_{-}-F_{a}F_{b}) + G_{bc}^2 F_a + G_{ac}^2 F_b + G_{bc}G_{bc}(M_{+}+M_{-})}\\
		r_2^{bb} &= 1 + \frac
		{i (F_aF_{c} - G_{ac}^2) \kappa_{c,2}}
		{F_c (M_{+}M_{-}-F_{a}F_{b}) + G_{bc}^2 F_a + G_{ac}^2 F_b + G_{bc}G_{bc}(M_{+}+M_{-})}\\
		r_2^{cc} &= 1 + \frac
		{i (F_aF_{b} - M_{+}M_{-}) \kappa_{c,3}}
		{F_c (M_{+}M_{-}-F_{a}F_{b}) + G_{bc}^2 F_a + G_{ac}^2 F_b + G_{bc}G_{bc}(M_{+}+M_{-})},
	\end{split}
\end{equation}
where $F_{\alpha (\alpha=a,b,c)}$ needs to be modified to $F_{\alpha} = \omega_{eff,\alpha} - 
\delta_{\alpha c}\omega_{eff,1}J_{\alpha}^{2}/D-g_{\alpha}^{2}/\omega_{eff,2}$ and $\omega_{eff,c}=\omega+i\frac{\kappa_{c,3}+\gamma_{c}}{2}$, while all other coefficients remain consistent with those in Sec. \ref{II}. 

\begin{figure}[htb]
	
	\includegraphics[width=15cm]{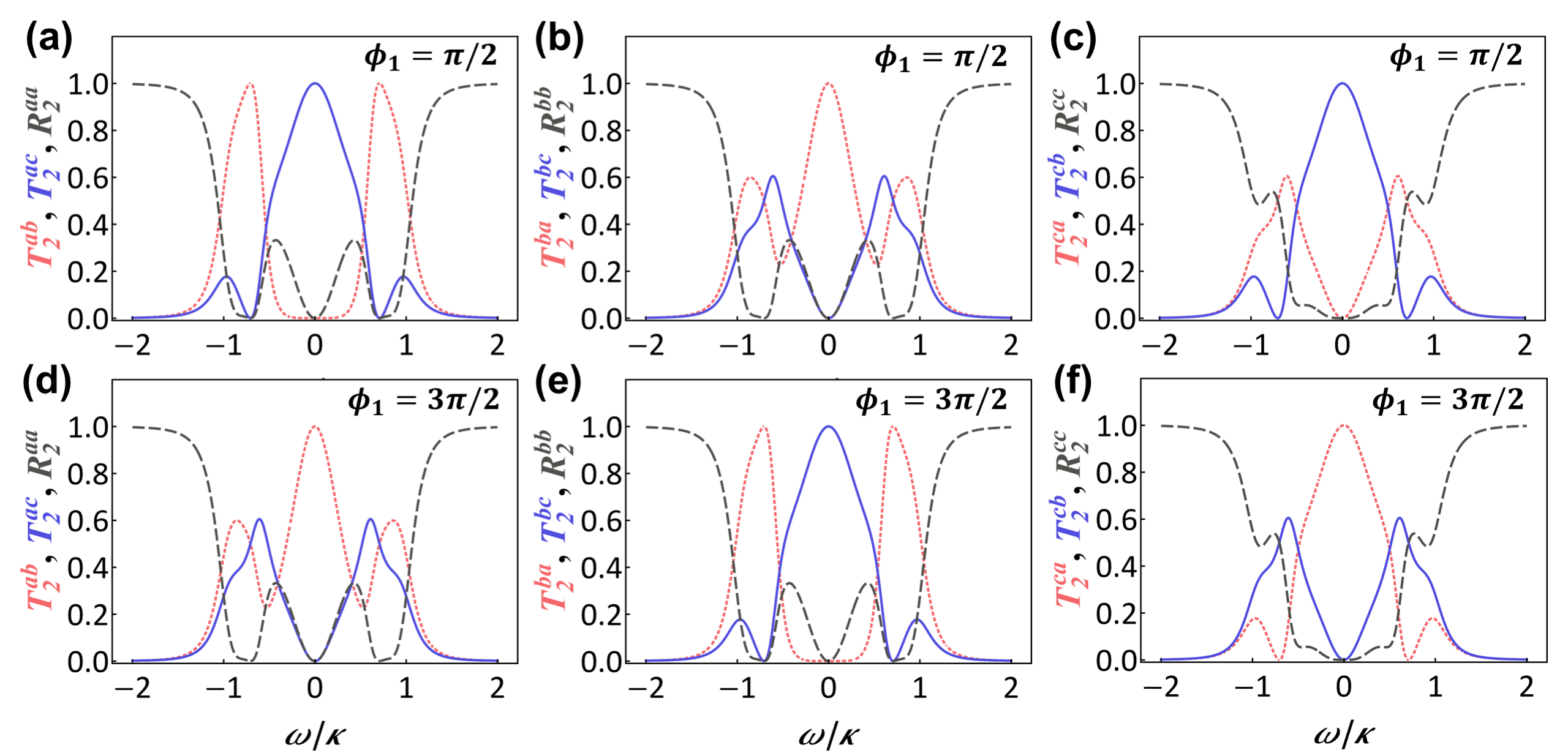}
	\centering
	
	\caption{Transmission $T_2^{\alpha \beta} $ and reflection probability $R_2^{\alpha \alpha}$ as functions of incident signal frequency $\omega/\kappa$ for different phase: (a), (b), and (c) $\phi_1=\pi/2$; (d), (e), and (f) $\phi_1=3\pi/2$. Figs. (a) and (d) represent the scattering at incidence from port a, while figs. (b) and (e), (c) and (f) represent the scattering at incidence from ports b, c, respectively. $g_a$, $g_b$, $J_{ab}$ and $g_c$ are given by Eq. \ref{eq23}, and the other parameters are set to $\kappa_c=\kappa$, $g=J_a=J_b=\kappa/2$, $\gamma_e=0$. 
	}
	
	\label{fig5}
\end{figure}

We begin by considering the optimal counterclockwise ($a \to b \to c$) circulator at $\omega = 0$, which prohibits signal transmission in the opposite direction. This implies that the condition $\ensuremath{\left|t_2^{ac}\right|}=\ensuremath{\left|t_2^{cb}\right|}=\ensuremath{\left|t_2^{ba}\right|}=0$ must always be satisfied. According to Eq. \ref{eq22}, those conditions imposed on the scattering matrix elements can be transformed into constraints on the parameters, i.e.,
\begin{equation}\label{eq23}
	\begin{split}
		g_a     &= g_b \equiv g\\
		J_{ab}  &= \frac{ 2J_a J_b}{\kappa_c}\\
		g_c^2   &= \frac{4g^2 - \gamma_e \kappa_c}{4}\\
		\phi_1  &= \frac{3\pi}{2},
	\end{split}
\end{equation}   

where $\kappa_{c,1} = \kappa_{c,2} = \kappa_{c,3} \equiv \kappa_c$ is set for simplicity. When the conditions in Eq. \ref{eq23} are applied to the expression of the scattering matrix elements in Eq. \ref{eq22}, the expression for the transmission probability $T_2^{\alpha \beta} \equiv \left|t_2^{\alpha \beta}\right|^{2} (\alpha,\beta=a,b,c)$ under unidirectional flow can be obtained by
\begin{equation}\label{eq24}
	\begin{split}
		T_2^{ab}         &=1\\
		T_2^{bc}=T_2^{ca}&=1-\frac{\gamma_e \kappa_c}{4 g^2}.
	\end{split}
\end{equation}  
It can be observed that when the condition $\gamma_e \kappa_c \ll 4 g^2$ is also satisfied, this system achieves optimal unidirectional circulation with counterclockwise direction at $\omega = 0 $.
It is worth noting that $T_2^{ab}=1$ always holds in this case. This is because the three-port circulator is equivalent to the two-port isolator shown in Fig. \ref{fig1}, but intrinsic damping rate of A-A is replaced by large damping rate induced by auxiliary cavity. Additionally, the condition for ensuring $\left| t_1^{ab} \right|=1$ (Eq. \ref{eq12}) does not involve damping rate of A-A and is also realized by Eq. \ref{eq23}. Therefore, $T_2^{ab}=1$ is valid in this context.
In Fig. \ref{fig5}, the transmission $T_2^{\alpha \beta} $ and reflection probability $R_2^{\alpha \alpha} \equiv \left|r_2^{\alpha \alpha}\right|^{2}$ as functions of the incident signal frequency $\omega / \kappa$ are shown for different phase $\phi_1$. The coupling parameters and damping rate are set to satisfy the constraints given in Eq. \ref{eq23} to ensure the absence of signal backflow in the opposite direction. When $\phi_1=\pi/2$, as shown in Figs. \ref{fig4}(a), \ref{fig4}(b), and \ref{fig4}(c), one can observe that at $\omega = 0$, $T_2^{ac}=T_2^{cb}=T_2^{ba}=1$, and all other scattering probabilities are zero. When $\phi_1=3\pi/2$, as shown in Figs. \ref{fig4}(d), \ref{fig4}(e), and \ref{fig4}(f), the system becomes $T_2^{ab}=T_2^{bc}=T_2^{ca}=1$, with all other scattering probabilities equal to zero. This means that when $\phi_1=\pi/2$, we can achieve a perfect circulator for a specific frequency signal, allowing transmission only in the clockwise direction. At the same time, we can conveniently switch to counterclockwise signal transmission by adjusting the externally applied magnetic flux to change the phase to $\phi_1=3\pi/2$. 

It is worth noting that, the intrinsic damping rate $\gamma_e$ of A-A plays an indispensable role in generating nonreciprocity for a two-port isolator, as discussed in Sec. \ref{III}. However, for the three-port circulator described here, when $\gamma_e = 0$, nonreciprocity still exists and even reaches optimal performance. This is because the additional damping rate induced by the coupling between the auxiliary cavity $c_2$ and A-A can effectively replace atom's intrinsic damping rate to generate nonreciprocity. This can be better understood by adiabatically eliminating the $c_2$ mode\cite{adiabatically_eliminating_1,adiabatically_eliminating_2,adiabatically_eliminating_3}.
Assuming that the external damping rate of the cavity mode $c_2$ is much greater than both the intrinsic damping rate $\gamma_e $ of the A-A and the coupling $g_c$ between the A-A and the cavity $c_2$ (i.e., $\kappa_{c,3} \gg \gamma_e, g_c$), one can adiabatically eliminate the $c_2$ mode. The QLEs Eq. \ref{eq3} remains unchanged, except that the equation of operator $\sigma$ is modified to the following form 
\begin{equation}\label{eq25}
	\dot{\sigma} = -i \left( g_{a} a_{2} + g_{b} b_{2} \right) - \frac{1}{2} \left( \gamma_{e} + \gamma_{e,id} + \kappa_{e,id} \right) \sigma -i \sqrt{\kappa_{e,id}} c_2^{in},
\end{equation}    
where $\gamma_{e,id} \equiv 4 g_c^2 \gamma_c/\left( \kappa_c,3 + \gamma_c \right)^2$ and $\kappa_{e,id} \equiv 4 g_c^2 \kappa_c,3/\left( \kappa_c,3 + \gamma_c \right)^2$ represent the effective damping rate of the A-A induced by the intrinsic and external damping rate of the auxiliary cavity mode $c_2$. It is evident that even if the intrinsic damping rate $\gamma_e $ of the atom vanishes, the total damping rate (i.e., $\gamma_e+ \gamma_{e,id}+\kappa_{e,id}$) still exists under the influence of the auxiliary cavity, ensuring the generation of nonreciprocity.

\begin{figure}[htb]
	
	\includegraphics[width=15cm]{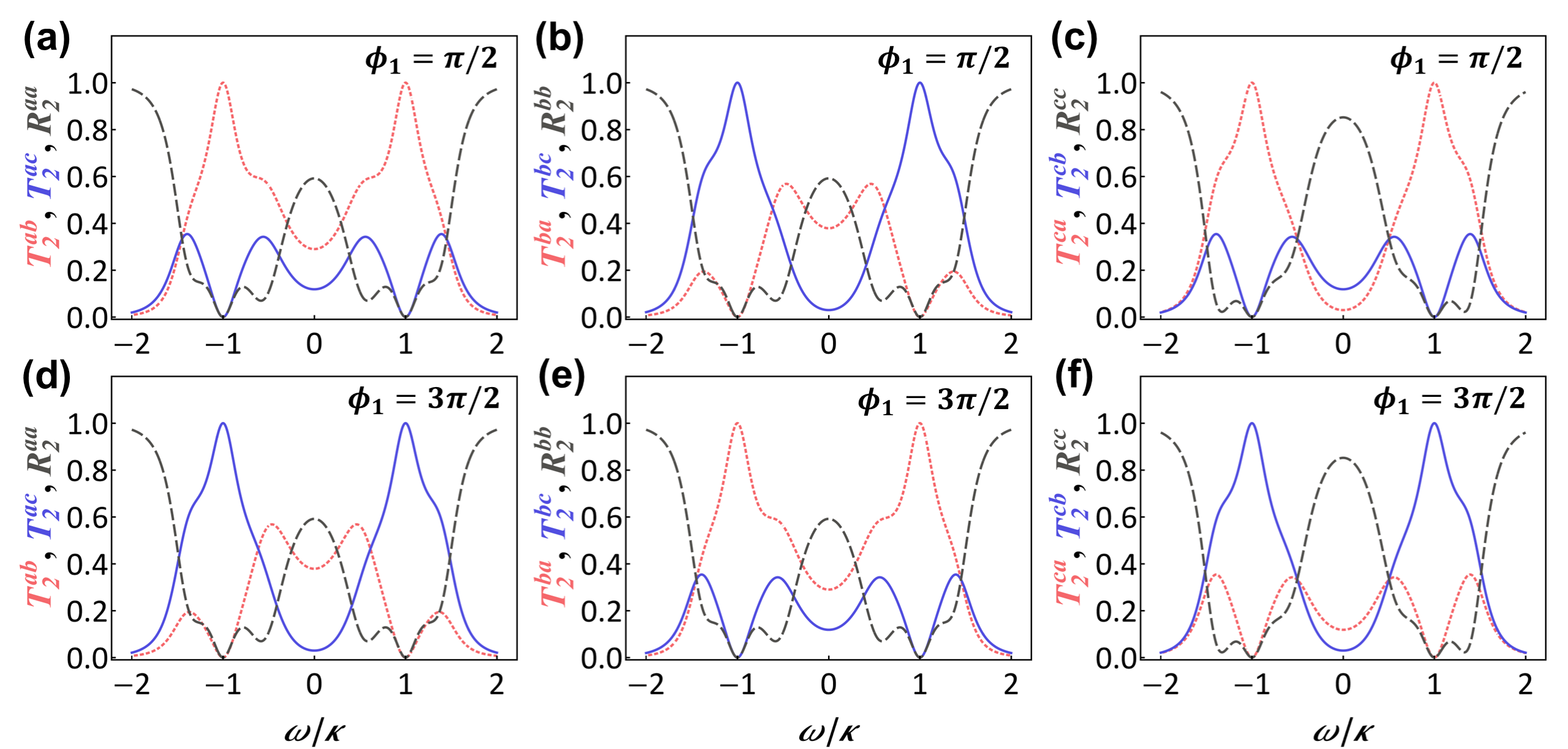}
	\centering
	
	\caption{Transmission $T_2^{\alpha \beta} $ and reflection probability $R_2^{\alpha \alpha}$ as functions of incident signal frequency $\omega/\kappa$ for different phase: (a), (b), and (c) $\phi_1=\pi/2$; (d), (e), and (f) $\phi_1=3\pi/2$. Figs. (a) and (d) represent the scattering at incidence from port a, while figs. (b) and (e), (c) and (f) represent the scattering at incidence from ports b, c, respectively. $g_a$, $g_b$, $J_{a}$, $J_{b}$, $J_{ab}$ and $g_c$ satisfy the conditions $g_a=g_b=\sqrt{\kappa_c(\kappa_c+\gamma_e)/4}$, $J_a=J_b=\sqrt{(4-\kappa_c^2)/4}$, $J_{ab}=\kappa_c/2$, $g_c=\sqrt{(4+\kappa_c^2)/4}$ to ensure unidirectional circulation in the system, and the other parameters are set to $\kappa_c=\kappa$, $\gamma_e=0$. 
	}
	
	\label{fig6}
\end{figure}

In the previous discussion of the three-port optical circulator shown in Fig. \ref{fig4}, the frequency of the signal generating the optimal nonreciprocal circulator was set at $\omega = 0$. In fact, through appropriate parameter adjustments, this frequency can be any point within the bandwidth where effective scattering occurs, and more complex scattering behaviors can emerge for other frequencies. When the parameters are adjusted to satisfy $g_a=g_b=\sqrt{\kappa_c(\kappa_c+\gamma_e)/4}$, $J_a=J_b=\sqrt{(4-\kappa_c^2)/4}$, $J_{ab}=\kappa_c/2$ and $g_c=\sqrt{(4+\kappa_c^2)/4}$, as shown in Fig. \ref{fig6}, we observe that at $\omega = \pm \kappa$, $T_2^{ab} = T_2^{bc} = T_2^{ca} = 1$ with all other scattering probabilities are zero, for phase $\phi_1=\pi/2$ (corresponding to Figs. \ref{fig6}(a), (b) and (c)). This means that perfect optical circulator behavior can be achieved for signals at two different frequencies, and due to the symmetry of the system, the circulation direction at both frequencies remains the same (counterclockwise). Similarly, by adjusting the externally applied magnetic flux to change the phase to $\phi_1=3\pi/2$, the direction of the optical symmetric-circulator at these two frequencies can be switched to clockwise, as shown in Figs. \ref{fig6}(d), (e) and (f).

\begin{figure}[htb]
	
	\centering
	\includegraphics[width=0.7\linewidth]{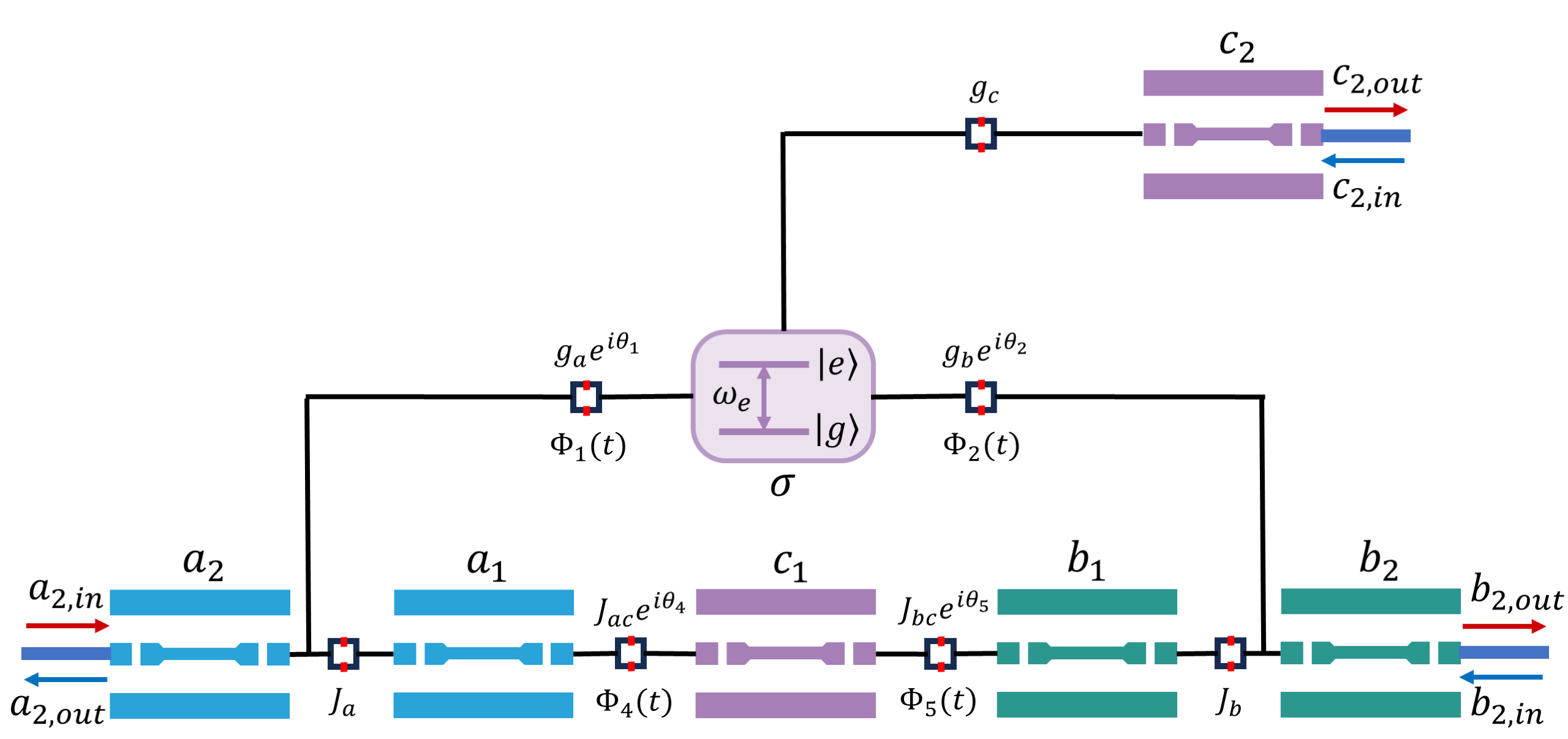}
	
	\caption{Schematic diagram of three-port antisymmetric-circulator with introducing an additional transition cavity $c_1$.}
	
	\label{fig7}
\end{figure}
If the goal is for the designed three-port circulator to exhibit opposite circulation directions at different frequencies, rather than the same direction shown in Fig. \ref{fig6}, an additional cavity can be introduced to modify the system’s symmetry. As shown in Fig. \ref{fig7}, a transition cavity $c_1$ resonant with the A-A, is used to couple the two cavities $a_1$ and $b_1$ with different frequencies. The Hamiltonian for this system is given by
\begin{equation}\label{eq26}
	\begin{split}
		\hat{H}_3  = \hat{H}_2 + \omega_ec_{1}^{\dagger}c_{1} + \left( J_{ac}(t)a_{1}^{\dagger}c_{1} + J_{bc}(t)b_{1}^{\dagger}c_{1} - J_{ab}(t)a_{1}^{\dagger}b_{1} + \text{\text{H.c.}} \right).
	\end{split}
\end{equation}    
Here, $J_{ac}(t)=2J_{ac}cos(\omega_{1}t+\theta_{4})$ ($J_{bc}(t)=2J_{bc}cos(\omega_{2}t+\theta_{5})$) is the time-dependent coupling between the cavity modes $a_1$ (or $b_1$) and the transition cavity $c_1$, generated by SQUIDs under an externally applied magnetic flux $\Phi_4(t)$ ($\Phi_5(t)$). The frequencies of the external magnetic flux are also chosen as $\omega_1 = \omega_e - \omega_a $ and $\omega_2 = \omega_e - \omega_b$, to match the detunings between different modes. In the interaction picture relative to $\hat{H}_{3,0}= \hat{H}_{2,0} + \omega_{e}c_1^{\dagger}c_1$, the Hamiltonian in Eq. \ref{eq26} can be written as
\begin{equation}\label{eq27}
	\hat{H}_{3,int} = \hat{H}_{2,int} + \left( J_{ac}a_{1}^{\dagger}c_{1} + J_{bc}e^{i\phi_2}b_{1}^{\dagger}c_{1} - J_{ab}e^{i\phi_1}a_{1}^{\dagger}b_{1} + \text{\text{H.c.}} \right),
\end{equation}
similarly, all phases $\theta_n$ in this system can be absorbed into phase $\phi_2$ ($\phi_2 = \theta_1 + \theta_5 - \theta_2 - \theta_4 $), and retained only in the terms $b_1^\dagger c_1$ and $c_1^\dagger b_1 $. Then the QLEs is given by
\begin{equation}\label{eq28}
	\begin{split}
		\dot{a}_{1} & = -i \left( J_{a} a_{2} + J_{ac} c_{1} \right) - \frac{1}{2} \gamma_{c} a_{1}\\
		\dot{a}_{2} & = -i \left( J_{a} a_{1} + g_{a} \sigma \right) - \frac{1}{2} \left( \kappa_{c,1} + \gamma_{c} \right) a_{2} + \sqrt{\kappa_{c,1}} a_{2,in}\\
		\dot{b}_{1} & = -i \left( J_{b} b_{2} + J_{bc} e^{i\phi_2} c_{1} \right) - \frac{1}{2} \gamma_{c} b_{1}\\
		\dot{b}_{2} & = -i \left( J_{b} b_{1} + g_{b} \sigma \right) - \frac{1}{2} \left( \kappa_{c,2} + \gamma_{c} \right) b_{2} + \sqrt{\kappa_{c,2}} b_{2,in}\\
		\dot{c}_{1} & = -i \left( J_{ab} a_{1} + J_{bc} e^{-i\phi_2} b_{1} \right) - \frac{1}{2} \gamma_{c} c_{1}\\
		\dot{c}_{2} & = -i g_{c} \sigma  - \frac{1}{2} \left( \kappa_{c,3} + \gamma_{c} \right) c_{2} + \sqrt{\kappa_{c,3}} c_{2,in}\\
		\dot{\sigma} & = -i \left( g_{a} a_{2} + g_{b} b_{2} + g_{c} c_{2} \right) - \frac{1}{2} \gamma_{e} \sigma. 
	\end{split}
\end{equation}

\begin{figure}[htb]
	
	\includegraphics[width=15cm]{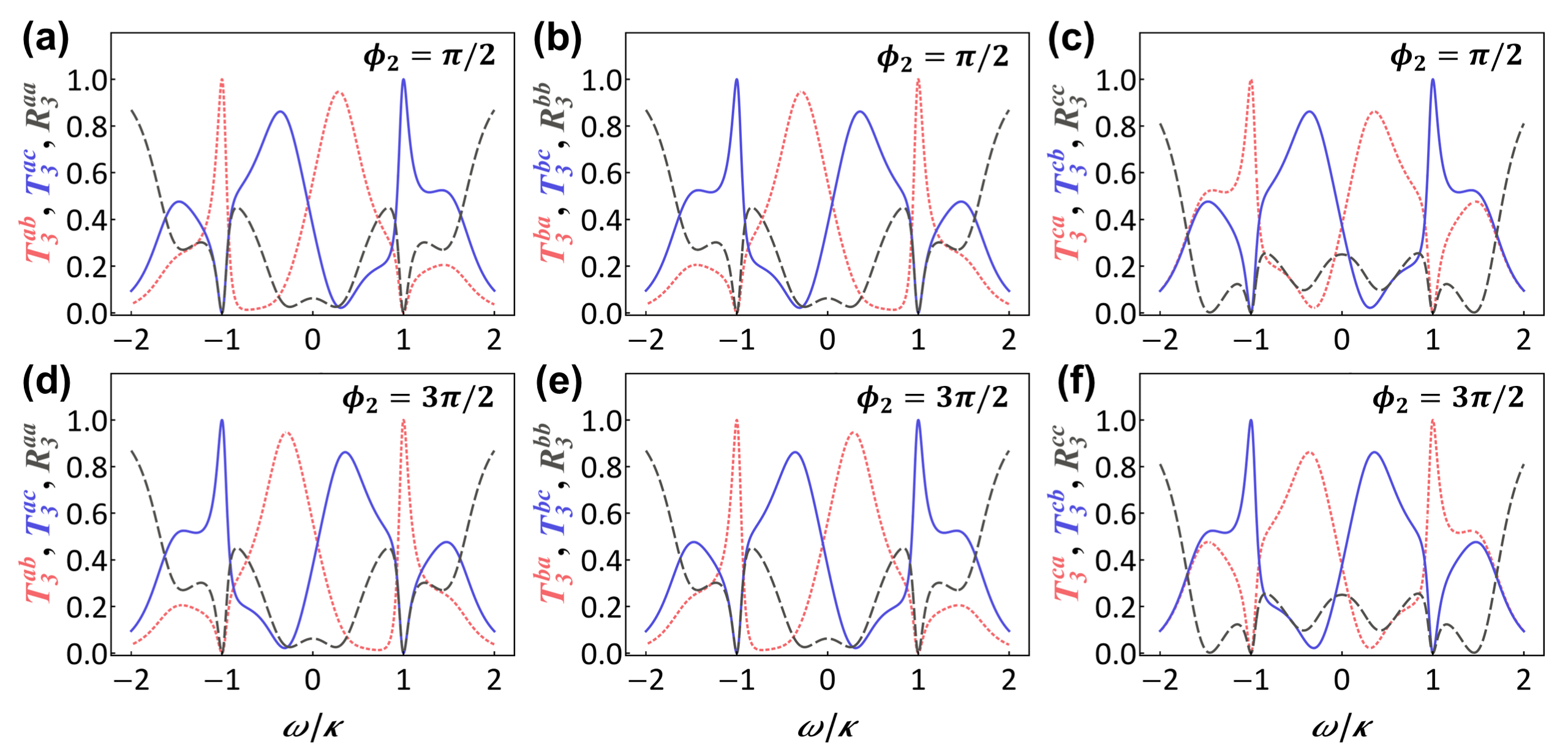}
	\centering
	
	\caption{Transmission probability $T_3^{\alpha \beta} $ and reflection probability $R_3^{\alpha \alpha}$ of the antisymmetric-circulator shown in Fig. \ref{fig7} as functions of the incident signal frequency $\omega/\kappa$ for different phase: (a), (b), and (c) $\phi_2=\pi/2$; (d), (e), and (f) $\phi_2=3\pi/2$. Figs. (a) and (d) represent the scattering at incidence from port a, while figs. (b) and (e), (c) and (f) represent the scattering at incidence from ports b, c, respectively. $g_a$, $g_b$, $J_{a}$, $J_{b}$, $J_{ac}$, $J_{bc}$ and $g_c$ satisfy the conditions $g_a=g_b=\sqrt{\kappa_c(\kappa_c+\gamma_e)/4}$, $J_a=J_b=\sqrt{(2-\kappa_c)/2}$, $J_{ac}=J_{bc}=\sqrt{\kappa_c/(2+\kappa_c)}$, $g_c=\sqrt{(4+\kappa_c^2)/4}$ to ensure unidirectional circulation in the system, and the other parameters are set to $\kappa_c=\sqrt{2}\kappa$, $\gamma_e=0$. 
	}
	
	\label{fig8}
\end{figure}

Following the same procedure as in Sec. IV, the scattering matrix for this three-port antisymmetric-circulator shown in Fig. \ref{fig7} can be derived as
\begin{equation}\label{eq29}
	S_{3}(\omega) = 
	\begin{bmatrix}
		r_3^{aa} & t_3^{ba} & t_3^{ca}\\
		t_3^{ab} & r_3^{bb} & t_3^{cb}\\
		t_3^{ac} & t_3^{bc} & r_3^{cc}
	\end{bmatrix},
\end{equation}
where the expressions of matrix elements have the same form as Eq. \ref{eq22} after considering the replacements
\begin{equation}\label{eq30}
	\begin{split}
		F_{\alpha} &\longrightarrow F_{\alpha (\alpha=a,b,c)}^{'} = \omega_{eff,\alpha} - \delta_{\alpha c}\frac{ J_{\alpha}^2 }{ \omega_{eff,1} } (1+\frac{J_{\alpha c}^2}{D^{'}}) - \frac{g_{\alpha}^2}{\omega_{eff,2}}\\
		M_{\pm}    &\longrightarrow M_{\pm}^{'}                   = \frac{ J_aJ_bJ_{ac}J_{bc}e^{\mp i\phi_2} }{ \omega_{eff,1}D^{'} } + \frac{g_ag_b}{\omega_{eff,2}},
	\end{split}
\end{equation}
with $D^{'} = \omega_{eff,1}^2 - J_{ac}^2 - J_{bc}^2$. Using the matrix elements expressions Eq. \ref{eq22} and Eq. \ref{eq30}, the transmission $T_3^{\alpha \beta}$ and reflection probability $R_3^{\alpha \alpha}$ of signal passing through this antisymmetric-circulator for different phase $\phi_2$ as functions of the frequency $\omega/\kappa$ are displayed in Fig. \ref{fig8}. It can be observed that when $\phi_2=\pi/2$, as shown in Figs \ref{fig8}(a), (b) and (c), at $\omega = -\kappa$, $t_3^{ab} = t_3^{bc} = t_3^{ca} = 1$, and all other scattering probabilities are zero. This corresponds to a circulator that allows signal transmission only in the counterclockwise direction ($a \to b \to c \to a$). Conversely, at $\omega = \kappa$, the situation is completely reversed (i.e., $t_3^{ac} = t_3^{cb} = t_3^{ba} = 1$, and all other scattering probabilities are zero), with the signal being transmitted only in the clockwise direction. 
This means that, for the antisymmetric-circulator, the direction of signal can be switched to the opposite direction as the incident signal frequency changes, while the symmetric-circulator above can only maintain the original direction. This change is mathematically attributed to the symmetry in the scattering probabilities caused by the introduction of the transition cavity $c_1$ with specific frequencie, i.e., $T_3^{\alpha\beta}(\omega) = T_3^{\beta\alpha}(-\omega)$. This symmetry in the symmetric-circulator system shown in Fig. \ref{fig5} is replaced by $T_3^{\alpha\beta}(\omega) = T_3^{\alpha\beta}(-\omega)$, so it maintains the same transmission direction for signals at both frequencies.

\section{Multifunctional Nonreciprocal Quantum Device}\label{V}

\begin{figure}[htb]
	
	\centering
	\includegraphics[width=0.7\linewidth]{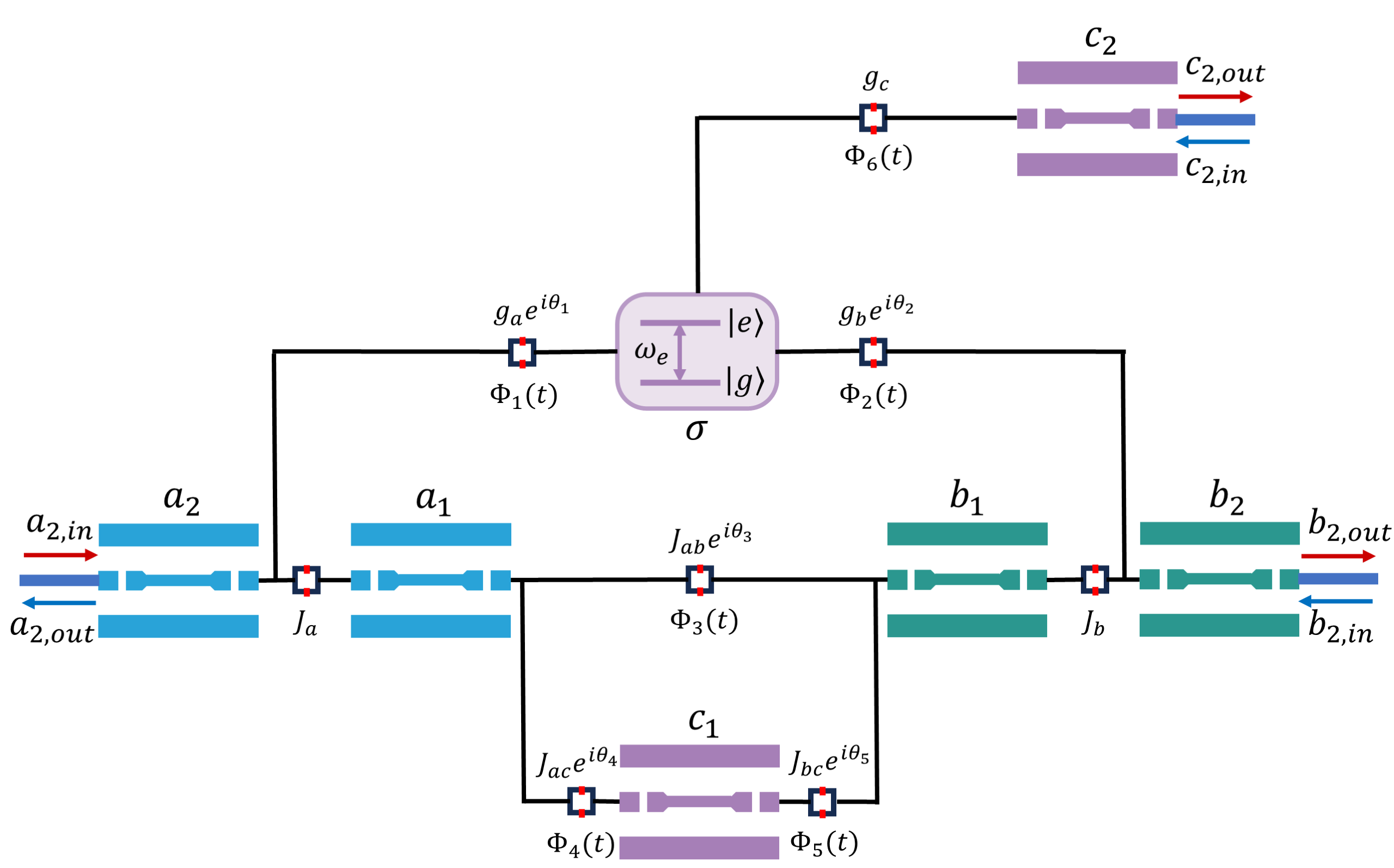}
	
	\caption{Schematic diagram of a multifunctional nonreciprocal quantum device. Overall adjustment of the externally applied magnetic flux $\Phi_n(t)$ can deactivate or restore interaction between different modes, thereby enabling the switching between two-port isolator, three-port symmetric-circulator, and antisymmetric-circulator.}
	
	\label{fig9}
\end{figure}

In fact, by utilizing the advantage of highly tunable interaction based on SQUID in SQC, the three devices with different functions described above can be integrated into the same superconducting architecture. As shown in Fig. \ref{fig9}, the multifunctional nonreciprocal quantum device can achieve three different nonreciprocal functions by overall adjusting the externally applied magnetic flux $\Phi_n(t)$. For example, by adjusting $\Phi_4(t)$, $\Phi_5(t)$, $\Phi_6(t)$ to static magnetic flux and setting appropriate values, the couplings between modes $a_1$ and $c_1$, $b_1$ and $c_1$, $c_2$ and $\sigma$ can be completely deactivated\cite{LCC_4,SQUID_1,selection_rule}, resulting in the two-port isolator shown in Fig. \ref{fig1}. Correspondingly, by deactivating the couplings between modes $a_1$ and $c_1$, $b_1$ and $c_1$, and restoring the coupling between $c_2$ and $\sigma$, the three-port symmetric-circulator shown in Fig. \ref{fig4} can be achieved. Finally, by deactivating the coupling between modes $a_1$ and $b_1$ (controlled by the external magnetic flux $\Phi_3(t)$), while activating the couplings between $a_1$ and $c_1$, $b_1$ and $c_1$, $c_2$ and $\sigma$, the three-port antisymmetric-circulator shown in Fig. \ref{fig7} can also be obtained. In other words, this multifunctional nonreciprocal device offers a wealth of expansion capabilities and operational convenience, as different functions can be switched according to actual needs simply by adjusting the externally applied magnetic flux.

\section{Conclusion}\label{VI}
In summary, we propose a multifunctional nonreciprocal quantum device based on SQC. It can achieve three nonreciprocal transmission functionalities under the control of external magnetic flux. First, it can function as a two-port isolator to protect fragile signals from harmful backflow noise. Our results demonstrate that this isolator can achieve nearly perfect unidirectional signal transmission while completely suppressing transmission in the opposite direction, when the intrinsic damping rates of various modes are considered. Additionally, the direction of signal transmission in the isolator can be conveniently controlled by the externally applied magnetic flux. Second, the device can implement two types of circulator functionalities. For the symmetric-circulator, the incident signal exhibits unidirectional circulation with same direction at two different frequencies, while for the antisymmetric-circulator, the signal circulates in the opposite direction. Similarly, the direction of signal transmission in the circulator can also be controlled by the externally applied magnetic flux.

The proposed multifunctional nonreciprocal quantum device currently supports only three functionalities. However, leveraging the advantages of integrability and tunable interaction of SQC, additional functionalities can be incorporated into this device following similar procedures. This significant potential for multifunctional integrated design could provide new insight for the development of large-scale quantum networks.

\section*{Data availability statement}
All data that support the findings of this study are included within the article (and any supplementary files).

\section*{ACKNOWLEDGMENTS}
This work was supported by National Natural Science Foundation of China (Grants No. 11874190 and No. 12247101).

\section*{Conflict of interest}
The authors declare no conflicts of interest.

\section*{ORCID iDs}
Lei Tan \orcidlink{0000-0002-9974-7766} \url{https://orcid.org/0000-0002-9974-7766}

\bibliographystyle{apsrev4-2}
\bibliography{refs}

\begin{thebibliography}{57}%
\makeatletter
\providecommand \@ifxundefined [1]{%
 \@ifx{#1\undefined}
}%
\providecommand \@ifnum [1]{%
 \ifnum #1\expandafter \@firstoftwo
 \else \expandafter \@secondoftwo
 \fi
}%
\providecommand \@ifx [1]{%
 \ifx #1\expandafter \@firstoftwo
 \else \expandafter \@secondoftwo
 \fi
}%
\providecommand \natexlab [1]{#1}%
\providecommand \enquote  [1]{``#1''}%
\providecommand \bibnamefont  [1]{#1}%
\providecommand \bibfnamefont [1]{#1}%
\providecommand \citenamefont [1]{#1}%
\providecommand \href@noop [0]{\@secondoftwo}%
\providecommand \href [0]{\begingroup \@sanitize@url \@href}%
\providecommand \@href[1]{\@@startlink{#1}\@@href}%
\providecommand \@@href[1]{\endgroup#1\@@endlink}%
\providecommand \@sanitize@url [0]{\catcode `\\12\catcode `\$12\catcode
  `\&12\catcode `\#12\catcode `\^12\catcode `\_12\catcode `\%12\relax}%
\providecommand \@@startlink[1]{}%
\providecommand \@@endlink[0]{}%
\providecommand \url  [0]{\begingroup\@sanitize@url \@url }%
\providecommand \@url [1]{\endgroup\@href {#1}{\urlprefix }}%
\providecommand \urlprefix  [0]{URL }%
\providecommand \Eprint [0]{\href }%
\providecommand \doibase [0]{https://doi.org/}%
\providecommand \selectlanguage [0]{\@gobble}%
\providecommand \bibinfo  [0]{\@secondoftwo}%
\providecommand \bibfield  [0]{\@secondoftwo}%
\providecommand \translation [1]{[#1]}%
\providecommand \BibitemOpen [0]{}%
\providecommand \bibitemStop [0]{}%
\providecommand \bibitemNoStop [0]{.\EOS\space}%
\providecommand \EOS [0]{\spacefactor3000\relax}%
\providecommand \BibitemShut  [1]{\csname bibitem#1\endcsname}%
\let\auto@bib@innerbib\@empty
\bibitem [{\citenamefont {Kimble}(2008)}]{quantum_network_1}%
  \BibitemOpen
  \bibfield  {author} {\bibinfo {author} {\bibfnamefont {H.~J.}\ \bibnamefont
  {Kimble}},\ }\bibfield  {title} {\bibinfo {title} {The quantum internet},\
  }\href {https://doi.org/10.1038/nature07127} {\bibfield  {journal} {\bibinfo
  {journal} {Nature}\ }\textbf {\bibinfo {volume} {453}},\ \bibinfo {pages}
  {1023} (\bibinfo {year} {2008})}\BibitemShut {NoStop}%
\bibitem [{\citenamefont {Wehner}\ \emph {et~al.}(2018)\citenamefont {Wehner},
  \citenamefont {Elkouss},\ and\ \citenamefont {Hanson}}]{quantum_network_2}%
  \BibitemOpen
  \bibfield  {author} {\bibinfo {author} {\bibfnamefont {S.}~\bibnamefont
  {Wehner}}, \bibinfo {author} {\bibfnamefont {D.}~\bibnamefont {Elkouss}},\
  and\ \bibinfo {author} {\bibfnamefont {R.}~\bibnamefont {Hanson}},\
  }\bibfield  {title} {\bibinfo {title} {Quantum internet: A vision for the
  road ahead},\ }\href {https://doi.org/10.1126/science.aam9288} {\bibfield
  {journal} {\bibinfo  {journal} {Science}\ }\textbf {\bibinfo {volume}
  {362}},\ \bibinfo {pages} {eaam9288} (\bibinfo {year} {2018})}\BibitemShut
  {NoStop}%
\bibitem [{\citenamefont {Chiribella}\ \emph {et~al.}(2009)\citenamefont
  {Chiribella}, \citenamefont {D'Ariano},\ and\ \citenamefont
  {Perinotti}}]{quantum_network_3}%
  \BibitemOpen
  \bibfield  {author} {\bibinfo {author} {\bibfnamefont {G.}~\bibnamefont
  {Chiribella}}, \bibinfo {author} {\bibfnamefont {G.~M.}\ \bibnamefont
  {D'Ariano}},\ and\ \bibinfo {author} {\bibfnamefont {P.}~\bibnamefont
  {Perinotti}},\ }\bibfield  {title} {\bibinfo {title} {Theoretical framework
  for quantum networks},\ }\href {https://doi.org/10.1103/PhysRevA.80.022339}
  {\bibfield  {journal} {\bibinfo  {journal} {Phys. Rev. A}\ }\textbf {\bibinfo
  {volume} {80}},\ \bibinfo {pages} {022339} (\bibinfo {year}
  {2009})}\BibitemShut {NoStop}%
\bibitem [{\citenamefont {Lo~Piparo}\ \emph {et~al.}(2020)\citenamefont
  {Lo~Piparo}, \citenamefont {Hanks}, \citenamefont {Nemoto},\ and\
  \citenamefont {Munro}}]{quantum_network_4}%
  \BibitemOpen
  \bibfield  {author} {\bibinfo {author} {\bibfnamefont {N.}~\bibnamefont
  {Lo~Piparo}}, \bibinfo {author} {\bibfnamefont {M.}~\bibnamefont {Hanks}},
  \bibinfo {author} {\bibfnamefont {K.}~\bibnamefont {Nemoto}},\ and\ \bibinfo
  {author} {\bibfnamefont {W.~J.}\ \bibnamefont {Munro}},\ }\bibfield  {title}
  {\bibinfo {title} {Aggregating quantum networks},\ }\href
  {https://doi.org/10.1103/PhysRevA.102.052613} {\bibfield  {journal} {\bibinfo
   {journal} {Phys. Rev. A}\ }\textbf {\bibinfo {volume} {102}},\ \bibinfo
  {pages} {052613} (\bibinfo {year} {2020})}\BibitemShut {NoStop}%
\bibitem [{\citenamefont {Chang}\ \emph {et~al.}(2018)\citenamefont {Chang},
  \citenamefont {Douglas}, \citenamefont {Gonz\'alez-Tudela}, \citenamefont
  {Hung},\ and\ \citenamefont {Kimble}}]{quantum_network_5}%
  \BibitemOpen
  \bibfield  {author} {\bibinfo {author} {\bibfnamefont {D.~E.}\ \bibnamefont
  {Chang}}, \bibinfo {author} {\bibfnamefont {J.~S.}\ \bibnamefont {Douglas}},
  \bibinfo {author} {\bibfnamefont {A.}~\bibnamefont {Gonz\'alez-Tudela}},
  \bibinfo {author} {\bibfnamefont {C.-L.}\ \bibnamefont {Hung}},\ and\
  \bibinfo {author} {\bibfnamefont {H.~J.}\ \bibnamefont {Kimble}},\ }\bibfield
   {title} {\bibinfo {title} {Colloquium: Quantum matter built from nanoscopic
  lattices of atoms and photons},\ }\href
  {https://doi.org/10.1103/RevModPhys.90.031002} {\bibfield  {journal}
  {\bibinfo  {journal} {Rev. Mod. Phys.}\ }\textbf {\bibinfo {volume} {90}},\
  \bibinfo {pages} {031002} (\bibinfo {year} {2018})}\BibitemShut {NoStop}%
\bibitem [{\citenamefont {Ye}\ and\ \citenamefont
  {Lu}(2022)}]{quantum_technology}%
  \BibitemOpen
  \bibfield  {author} {\bibinfo {author} {\bibfnamefont {Z.}~\bibnamefont
  {Ye}}\ and\ \bibinfo {author} {\bibfnamefont {Y.}~\bibnamefont {Lu}},\
  }\bibfield  {title} {\bibinfo {title} {Quantum science: a review and current
  research trends},\ }\href {https://doi.org/10.1080/23270012.2022.2089064}
  {\bibfield  {journal} {\bibinfo  {journal} {Journal of Management Analytics}\
  }\textbf {\bibinfo {volume} {9}},\ \bibinfo {pages} {383} (\bibinfo {year}
  {2022})},\ \Eprint
  {https://arxiv.org/abs/https://doi.org/10.1080/23270012.2022.2089064}
  {https://doi.org/10.1080/23270012.2022.2089064} \BibitemShut {NoStop}%
\bibitem [{\citenamefont {Duan}\ and\ \citenamefont
  {Raussendorf}(2005)}]{quantum_computing_1}%
  \BibitemOpen
  \bibfield  {author} {\bibinfo {author} {\bibfnamefont {L.-M.}\ \bibnamefont
  {Duan}}\ and\ \bibinfo {author} {\bibfnamefont {R.}~\bibnamefont
  {Raussendorf}},\ }\bibfield  {title} {\bibinfo {title} {Efficient quantum
  computation with probabilistic quantum gates},\ }\href
  {https://doi.org/10.1103/PhysRevLett.95.080503} {\bibfield  {journal}
  {\bibinfo  {journal} {Phys. Rev. Lett.}\ }\textbf {\bibinfo {volume} {95}},\
  \bibinfo {pages} {080503} (\bibinfo {year} {2005})}\BibitemShut {NoStop}%
\bibitem [{\citenamefont {Zoller}\ \emph {et~al.}(2005)\citenamefont {Zoller},
  \citenamefont {Beth}, \citenamefont {Binosi}, \citenamefont {Blatt},
  \citenamefont {Briegel}, \citenamefont {Bruss}, \citenamefont {Calarco},
  \citenamefont {Cirac}, \citenamefont {Deutsch}, \citenamefont {Eisert},
  \citenamefont {Ekert}, \citenamefont {Fabre}, \citenamefont {Gisin},
  \citenamefont {Grangiere}, \citenamefont {Grassl}, \citenamefont {Haroche},
  \citenamefont {Imamoglu}, \citenamefont {Karlson}, \citenamefont {Kempe},
  \citenamefont {Kouwenhoven}, \citenamefont {Kr{\"o}ll}, \citenamefont
  {Leuchs}, \citenamefont {Lewenstein}, \citenamefont {Loss}, \citenamefont
  {L{\"u}tkenhaus}, \citenamefont {Massar}, \citenamefont {Mooij},
  \citenamefont {Plenio}, \citenamefont {Polzik}, \citenamefont {Popescu},
  \citenamefont {Rempe}, \citenamefont {Sergienko}, \citenamefont {Suter},
  \citenamefont {Twamley}, \citenamefont {Wendin}, \citenamefont {Werner},
  \citenamefont {Winter}, \citenamefont {Wrachtrup},\ and\ \citenamefont
  {Zeilinger}}]{quantum_communication_1}%
  \BibitemOpen
  \bibfield  {author} {\bibinfo {author} {\bibfnamefont {P.}~\bibnamefont
  {Zoller}}, \bibinfo {author} {\bibfnamefont {T.}~\bibnamefont {Beth}},
  \bibinfo {author} {\bibfnamefont {D.}~\bibnamefont {Binosi}}, \bibinfo
  {author} {\bibfnamefont {R.}~\bibnamefont {Blatt}}, \bibinfo {author}
  {\bibfnamefont {H.}~\bibnamefont {Briegel}}, \bibinfo {author} {\bibfnamefont
  {D.}~\bibnamefont {Bruss}}, \bibinfo {author} {\bibfnamefont
  {T.}~\bibnamefont {Calarco}}, \bibinfo {author} {\bibfnamefont {J.~I.}\
  \bibnamefont {Cirac}}, \bibinfo {author} {\bibfnamefont {D.}~\bibnamefont
  {Deutsch}}, \bibinfo {author} {\bibfnamefont {J.}~\bibnamefont {Eisert}},
  \bibinfo {author} {\bibfnamefont {A.}~\bibnamefont {Ekert}}, \bibinfo
  {author} {\bibfnamefont {C.}~\bibnamefont {Fabre}}, \bibinfo {author}
  {\bibfnamefont {N.}~\bibnamefont {Gisin}}, \bibinfo {author} {\bibfnamefont
  {P.}~\bibnamefont {Grangiere}}, \bibinfo {author} {\bibfnamefont
  {M.}~\bibnamefont {Grassl}}, \bibinfo {author} {\bibfnamefont
  {S.}~\bibnamefont {Haroche}}, \bibinfo {author} {\bibfnamefont
  {A.}~\bibnamefont {Imamoglu}}, \bibinfo {author} {\bibfnamefont
  {A.}~\bibnamefont {Karlson}}, \bibinfo {author} {\bibfnamefont
  {J.}~\bibnamefont {Kempe}}, \bibinfo {author} {\bibfnamefont
  {L.}~\bibnamefont {Kouwenhoven}}, \bibinfo {author} {\bibfnamefont
  {S.}~\bibnamefont {Kr{\"o}ll}}, \bibinfo {author} {\bibfnamefont
  {G.}~\bibnamefont {Leuchs}}, \bibinfo {author} {\bibfnamefont
  {M.}~\bibnamefont {Lewenstein}}, \bibinfo {author} {\bibfnamefont
  {D.}~\bibnamefont {Loss}}, \bibinfo {author} {\bibfnamefont {N.}~\bibnamefont
  {L{\"u}tkenhaus}}, \bibinfo {author} {\bibfnamefont {S.}~\bibnamefont
  {Massar}}, \bibinfo {author} {\bibfnamefont {J.~E.}\ \bibnamefont {Mooij}},
  \bibinfo {author} {\bibfnamefont {M.~B.}\ \bibnamefont {Plenio}}, \bibinfo
  {author} {\bibfnamefont {E.}~\bibnamefont {Polzik}}, \bibinfo {author}
  {\bibfnamefont {S.}~\bibnamefont {Popescu}}, \bibinfo {author} {\bibfnamefont
  {G.}~\bibnamefont {Rempe}}, \bibinfo {author} {\bibfnamefont
  {A.}~\bibnamefont {Sergienko}}, \bibinfo {author} {\bibfnamefont
  {D.}~\bibnamefont {Suter}}, \bibinfo {author} {\bibfnamefont
  {J.}~\bibnamefont {Twamley}}, \bibinfo {author} {\bibfnamefont
  {G.}~\bibnamefont {Wendin}}, \bibinfo {author} {\bibfnamefont
  {R.}~\bibnamefont {Werner}}, \bibinfo {author} {\bibfnamefont
  {A.}~\bibnamefont {Winter}}, \bibinfo {author} {\bibfnamefont
  {J.}~\bibnamefont {Wrachtrup}},\ and\ \bibinfo {author} {\bibfnamefont
  {A.}~\bibnamefont {Zeilinger}},\ }\bibfield  {title} {\bibinfo {title}
  {Quantum information processing and communication},\ }\href
  {https://doi.org/10.1140/epjd/e2005-00251-1} {\bibfield  {journal} {\bibinfo
  {journal} {Eur. Phys. J. D - Atomic, Molecular, Optical and Plasma Physics}\
  }\textbf {\bibinfo {volume} {36}},\ \bibinfo {pages} {203} (\bibinfo {year}
  {2005})}\BibitemShut {NoStop}%
\bibitem [{\citenamefont {Bennett}\ and\ \citenamefont
  {Brassard}(2014)}]{quantum_communication_2}%
  \BibitemOpen
  \bibfield  {author} {\bibinfo {author} {\bibfnamefont {C.~H.}\ \bibnamefont
  {Bennett}}\ and\ \bibinfo {author} {\bibfnamefont {G.}~\bibnamefont
  {Brassard}},\ }\bibfield  {title} {\bibinfo {title} {Quantum cryptography:
  Public key distribution and coin tossing},\ }\href
  {https://doi.org/10.1016/j.tcs.2014.05.025} {\bibfield  {journal} {\bibinfo
  {journal} {Theor Comput Sci}\ }\textbf {\bibinfo {volume} {560}},\ \bibinfo
  {pages} {7} (\bibinfo {year} {2014})},\ \bibinfo {note} {theoretical Aspects
  of Quantum Cryptography – celebrating 30 years of BB84}\BibitemShut
  {NoStop}%
\bibitem [{\citenamefont {Qiu}(2014)}]{quantum_communication_3}%
  \BibitemOpen
  \bibfield  {author} {\bibinfo {author} {\bibfnamefont {J.}~\bibnamefont
  {Qiu}},\ }\bibfield  {title} {\bibinfo {title} {Quantum communications leap
  out of the lab},\ }\href {https://doi.org/10.1038/508441a} {\bibfield
  {journal} {\bibinfo  {journal} {Nature}\ }\textbf {\bibinfo {volume} {508}},\
  \bibinfo {pages} {441} (\bibinfo {year} {2014})}\BibitemShut {NoStop}%
\bibitem [{\citenamefont {Bennett}\ and\ \citenamefont
  {DiVincenzo}(2000)}]{quantum_communication_4}%
  \BibitemOpen
  \bibfield  {author} {\bibinfo {author} {\bibfnamefont {C.~H.}\ \bibnamefont
  {Bennett}}\ and\ \bibinfo {author} {\bibfnamefont {D.~P.}\ \bibnamefont
  {DiVincenzo}},\ }\bibfield  {title} {\bibinfo {title} {Quantum information
  and computation},\ }\href {https://doi.org/10.1038/35005001} {\bibfield
  {journal} {\bibinfo  {journal} {Nature}\ }\textbf {\bibinfo {volume} {404}},\
  \bibinfo {pages} {247} (\bibinfo {year} {2000})}\BibitemShut {NoStop}%
\bibitem [{\citenamefont {{Gallego Torromé}}\ and\ \citenamefont
  {Barzanjeh}(2024)}]{quantum_radar_1}%
  \BibitemOpen
  \bibfield  {author} {\bibinfo {author} {\bibfnamefont {R.}~\bibnamefont
  {{Gallego Torromé}}}\ and\ \bibinfo {author} {\bibfnamefont
  {S.}~\bibnamefont {Barzanjeh}},\ }\bibfield  {title} {\bibinfo {title}
  {Advances in quantum radar and quantum lidar},\ }\href
  {https://doi.org/https://doi.org/10.1016/j.pquantelec.2023.100497} {\bibfield
   {journal} {\bibinfo  {journal} {Progress in Quantum Electronics}\ }\textbf
  {\bibinfo {volume} {93}},\ \bibinfo {pages} {100497} (\bibinfo {year}
  {2024})}\BibitemShut {NoStop}%
\bibitem [{\citenamefont {Assouly}\ \emph {et~al.}(2023)\citenamefont
  {Assouly}, \citenamefont {Dassonneville}, \citenamefont {Peronnin},
  \citenamefont {Bienfait},\ and\ \citenamefont {Huard}}]{quantum_radar_2}%
  \BibitemOpen
  \bibfield  {author} {\bibinfo {author} {\bibfnamefont {R.}~\bibnamefont
  {Assouly}}, \bibinfo {author} {\bibfnamefont {R.}~\bibnamefont
  {Dassonneville}}, \bibinfo {author} {\bibfnamefont {T.}~\bibnamefont
  {Peronnin}}, \bibinfo {author} {\bibfnamefont {A.}~\bibnamefont {Bienfait}},\
  and\ \bibinfo {author} {\bibfnamefont {B.}~\bibnamefont {Huard}},\ }\bibfield
   {title} {\bibinfo {title} {Quantum advantage in microwave quantum radar},\
  }\href {https://doi.org/10.1038/s41567-023-02113-4} {\bibfield  {journal}
  {\bibinfo  {journal} {Nature Physics}\ }\textbf {\bibinfo {volume} {19}},\
  \bibinfo {pages} {1418} (\bibinfo {year} {2023})}\BibitemShut {NoStop}%
\bibitem [{\citenamefont {Slepyan}\ \emph {et~al.}(2022)\citenamefont
  {Slepyan}, \citenamefont {Vlasenko}, \citenamefont {Mogilevtsev},\ and\
  \citenamefont {Boag}}]{quantum_radar_3}%
  \BibitemOpen
  \bibfield  {author} {\bibinfo {author} {\bibfnamefont {G.}~\bibnamefont
  {Slepyan}}, \bibinfo {author} {\bibfnamefont {S.}~\bibnamefont {Vlasenko}},
  \bibinfo {author} {\bibfnamefont {D.}~\bibnamefont {Mogilevtsev}},\ and\
  \bibinfo {author} {\bibfnamefont {A.}~\bibnamefont {Boag}},\ }\bibfield
  {title} {\bibinfo {title} {Quantum radars and lidars: Concepts, realizations,
  and perspectives},\ }\href {https://doi.org/10.1109/MAP.2021.3089994}
  {\bibfield  {journal} {\bibinfo  {journal} {IEEE Antennas and Propagation
  Magazine}\ }\textbf {\bibinfo {volume} {64}},\ \bibinfo {pages} {16}
  (\bibinfo {year} {2022})}\BibitemShut {NoStop}%
\bibitem [{\citenamefont {Zhu}(2022)}]{quantum_measurement_1}%
  \BibitemOpen
  \bibfield  {author} {\bibinfo {author} {\bibfnamefont {H.}~\bibnamefont
  {Zhu}},\ }\bibfield  {title} {\bibinfo {title} {Quantum measurements in the
  light of quantum state estimation},\ }\href
  {https://doi.org/10.1103/PRXQuantum.3.030306} {\bibfield  {journal} {\bibinfo
   {journal} {PRX Quantum}\ }\textbf {\bibinfo {volume} {3}},\ \bibinfo {pages}
  {030306} (\bibinfo {year} {2022})}\BibitemShut {NoStop}%
\bibitem [{\citenamefont {Kechrimparis}\ \emph {et~al.}(2020)\citenamefont
  {Kechrimparis}, \citenamefont {Kropf}, \citenamefont {Wudarski},\ and\
  \citenamefont {Bae}}]{quantum_measurement_2}%
  \BibitemOpen
  \bibfield  {author} {\bibinfo {author} {\bibfnamefont {S.}~\bibnamefont
  {Kechrimparis}}, \bibinfo {author} {\bibfnamefont {C.~M.}\ \bibnamefont
  {Kropf}}, \bibinfo {author} {\bibfnamefont {F.}~\bibnamefont {Wudarski}},\
  and\ \bibinfo {author} {\bibfnamefont {J.}~\bibnamefont {Bae}},\ }\bibfield
  {title} {\bibinfo {title} {Channel coding of a quantum measurement},\ }\href
  {https://doi.org/10.1109/JSAC.2020.2969034} {\bibfield  {journal} {\bibinfo
  {journal} {IEEE Journal on Selected Areas in Communications}\ }\textbf
  {\bibinfo {volume} {38}},\ \bibinfo {pages} {439} (\bibinfo {year}
  {2020})}\BibitemShut {NoStop}%
\bibitem [{\citenamefont {Zhang}\ \emph {et~al.}(2020)\citenamefont {Zhang},
  \citenamefont {Xie}, \citenamefont {Xu}, \citenamefont {Zheng}, \citenamefont
  {Zhang}, \citenamefont {Poon}, \citenamefont {Vedral},\ and\ \citenamefont
  {Zhang}}]{quantum_measurement_3}%
  \BibitemOpen
  \bibfield  {author} {\bibinfo {author} {\bibfnamefont {A.}~\bibnamefont
  {Zhang}}, \bibinfo {author} {\bibfnamefont {J.}~\bibnamefont {Xie}}, \bibinfo
  {author} {\bibfnamefont {H.}~\bibnamefont {Xu}}, \bibinfo {author}
  {\bibfnamefont {K.}~\bibnamefont {Zheng}}, \bibinfo {author} {\bibfnamefont
  {H.}~\bibnamefont {Zhang}}, \bibinfo {author} {\bibfnamefont {Y.-T.}\
  \bibnamefont {Poon}}, \bibinfo {author} {\bibfnamefont {V.}~\bibnamefont
  {Vedral}},\ and\ \bibinfo {author} {\bibfnamefont {L.}~\bibnamefont
  {Zhang}},\ }\bibfield  {title} {\bibinfo {title} {Experimental
  self-characterization of quantum measurements},\ }\href
  {https://doi.org/10.1103/PhysRevLett.124.040402} {\bibfield  {journal}
  {\bibinfo  {journal} {Phys. Rev. Lett.}\ }\textbf {\bibinfo {volume} {124}},\
  \bibinfo {pages} {040402} (\bibinfo {year} {2020})}\BibitemShut {NoStop}%
\bibitem [{\citenamefont {Jalas}\ \emph {et~al.}(2013)\citenamefont {Jalas},
  \citenamefont {Petrov}, \citenamefont {Eich}, \citenamefont {Freude},
  \citenamefont {Fan}, \citenamefont {Yu}, \citenamefont {Baets}, \citenamefont
  {Popovi{\'{c}}}, \citenamefont {Melloni}, \citenamefont {Joannopoulos},
  \citenamefont {Vanwolleghem}, \citenamefont {Doerr},\ and\ \citenamefont
  {Renner}}]{isolator_1}%
  \BibitemOpen
  \bibfield  {author} {\bibinfo {author} {\bibfnamefont {D.}~\bibnamefont
  {Jalas}}, \bibinfo {author} {\bibfnamefont {A.}~\bibnamefont {Petrov}},
  \bibinfo {author} {\bibfnamefont {M.}~\bibnamefont {Eich}}, \bibinfo {author}
  {\bibfnamefont {W.}~\bibnamefont {Freude}}, \bibinfo {author} {\bibfnamefont
  {S.}~\bibnamefont {Fan}}, \bibinfo {author} {\bibfnamefont {Z.}~\bibnamefont
  {Yu}}, \bibinfo {author} {\bibfnamefont {R.}~\bibnamefont {Baets}}, \bibinfo
  {author} {\bibfnamefont {M.}~\bibnamefont {Popovi{\'{c}}}}, \bibinfo {author}
  {\bibfnamefont {A.}~\bibnamefont {Melloni}}, \bibinfo {author} {\bibfnamefont
  {J.~D.}\ \bibnamefont {Joannopoulos}}, \bibinfo {author} {\bibfnamefont
  {M.}~\bibnamefont {Vanwolleghem}}, \bibinfo {author} {\bibfnamefont {C.~R.}\
  \bibnamefont {Doerr}},\ and\ \bibinfo {author} {\bibfnamefont
  {H.}~\bibnamefont {Renner}},\ }\bibfield  {title} {\bibinfo {title} {What is
  --- and what is not --- an optical isolator},\ }\href
  {https://doi.org/10.1038/nphoton.2013.185} {\bibfield  {journal} {\bibinfo
  {journal} {Nature Photonics}\ }\textbf {\bibinfo {volume} {7}},\ \bibinfo
  {pages} {579} (\bibinfo {year} {2013})}\BibitemShut {NoStop}%
\bibitem [{\citenamefont {Scheucher}\ \emph {et~al.}(2016)\citenamefont
  {Scheucher}, \citenamefont {Hilico}, \citenamefont {Will}, \citenamefont
  {Volz},\ and\ \citenamefont {Rauschenbeutel}}]{circulator_1}%
  \BibitemOpen
  \bibfield  {author} {\bibinfo {author} {\bibfnamefont {M.}~\bibnamefont
  {Scheucher}}, \bibinfo {author} {\bibfnamefont {A.}~\bibnamefont {Hilico}},
  \bibinfo {author} {\bibfnamefont {E.}~\bibnamefont {Will}}, \bibinfo {author}
  {\bibfnamefont {J.}~\bibnamefont {Volz}},\ and\ \bibinfo {author}
  {\bibfnamefont {A.}~\bibnamefont {Rauschenbeutel}},\ }\bibfield  {title}
  {\bibinfo {title} {Quantum optical circulator controlled by a single chirally
  coupled atom},\ }\href {https://doi.org/10.1126/science.aaj2118} {\bibfield
  {journal} {\bibinfo  {journal} {Science}\ }\textbf {\bibinfo {volume}
  {354}},\ \bibinfo {pages} {1577} (\bibinfo {year} {2016})},\ \Eprint
  {https://arxiv.org/abs/https://www.science.org/doi/pdf/10.1126/science.aaj2118}
  {https://www.science.org/doi/pdf/10.1126/science.aaj2118} \BibitemShut
  {NoStop}%
\bibitem [{\citenamefont {Haldane}\ and\ \citenamefont
  {Raghu}(2008)}]{symmetry_breaking_1}%
  \BibitemOpen
  \bibfield  {author} {\bibinfo {author} {\bibfnamefont {F.~D.~M.}\
  \bibnamefont {Haldane}}\ and\ \bibinfo {author} {\bibfnamefont
  {S.}~\bibnamefont {Raghu}},\ }\bibfield  {title} {\bibinfo {title} {Possible
  realization of directional optical waveguides in photonic crystals with
  broken time-reversal symmetry},\ }\href
  {https://doi.org/10.1103/PhysRevLett.100.013904} {\bibfield  {journal}
  {\bibinfo  {journal} {Phys. Rev. Lett.}\ }\textbf {\bibinfo {volume} {100}},\
  \bibinfo {pages} {013904} (\bibinfo {year} {2008})}\BibitemShut {NoStop}%
\bibitem [{\citenamefont {Caloz}\ \emph {et~al.}(2018)\citenamefont {Caloz},
  \citenamefont {Al\`u}, \citenamefont {Tretyakov}, \citenamefont {Sounas},
  \citenamefont {Achouri},\ and\ \citenamefont
  {Deck-L\'eger}}]{symmetry_breaking_2}%
  \BibitemOpen
  \bibfield  {author} {\bibinfo {author} {\bibfnamefont {C.}~\bibnamefont
  {Caloz}}, \bibinfo {author} {\bibfnamefont {A.}~\bibnamefont {Al\`u}},
  \bibinfo {author} {\bibfnamefont {S.}~\bibnamefont {Tretyakov}}, \bibinfo
  {author} {\bibfnamefont {D.}~\bibnamefont {Sounas}}, \bibinfo {author}
  {\bibfnamefont {K.}~\bibnamefont {Achouri}},\ and\ \bibinfo {author}
  {\bibfnamefont {Z.-L.}\ \bibnamefont {Deck-L\'eger}},\ }\bibfield  {title}
  {\bibinfo {title} {Electromagnetic nonreciprocity},\ }\href
  {https://doi.org/10.1103/PhysRevApplied.10.047001} {\bibfield  {journal}
  {\bibinfo  {journal} {Phys. Rev. Appl.}\ }\textbf {\bibinfo {volume} {10}},\
  \bibinfo {pages} {047001} (\bibinfo {year} {2018})}\BibitemShut {NoStop}%
\bibitem [{\citenamefont {Lodahl}\ \emph {et~al.}(2017)\citenamefont {Lodahl},
  \citenamefont {Mahmoodian}, \citenamefont {Stobbe}, \citenamefont
  {Rauschenbeutel}, \citenamefont {Schneeweiss}, \citenamefont {Volz},
  \citenamefont {Pichler},\ and\ \citenamefont {Zoller}}]{symmetry_breaking_3}%
  \BibitemOpen
  \bibfield  {author} {\bibinfo {author} {\bibfnamefont {P.}~\bibnamefont
  {Lodahl}}, \bibinfo {author} {\bibfnamefont {S.}~\bibnamefont {Mahmoodian}},
  \bibinfo {author} {\bibfnamefont {S.}~\bibnamefont {Stobbe}}, \bibinfo
  {author} {\bibfnamefont {A.}~\bibnamefont {Rauschenbeutel}}, \bibinfo
  {author} {\bibfnamefont {P.}~\bibnamefont {Schneeweiss}}, \bibinfo {author}
  {\bibfnamefont {J.}~\bibnamefont {Volz}}, \bibinfo {author} {\bibfnamefont
  {H.}~\bibnamefont {Pichler}},\ and\ \bibinfo {author} {\bibfnamefont
  {P.}~\bibnamefont {Zoller}},\ }\bibfield  {title} {\bibinfo {title} {Chiral
  quantum optics},\ }\href {https://doi.org/10.1038/nature21037} {\bibfield
  {journal} {\bibinfo  {journal} {Nature}\ }\textbf {\bibinfo {volume} {541}},\
  \bibinfo {pages} {473} (\bibinfo {year} {2017})}\BibitemShut {NoStop}%
\bibitem [{\citenamefont {Fay}\ and\ \citenamefont
  {Comstock}(1965)}]{ferrite_1}%
  \BibitemOpen
  \bibfield  {author} {\bibinfo {author} {\bibfnamefont {C.}~\bibnamefont
  {Fay}}\ and\ \bibinfo {author} {\bibfnamefont {R.}~\bibnamefont {Comstock}},\
  }\bibfield  {title} {\bibinfo {title} {Operation of the ferrite junction
  circulator},\ }\href {https://doi.org/10.1109/TMTT.1965.1125923} {\bibfield
  {journal} {\bibinfo  {journal} {IEEE Transactions on Microwave Theory and
  Techniques}\ }\textbf {\bibinfo {volume} {13}},\ \bibinfo {pages} {15}
  (\bibinfo {year} {1965})}\BibitemShut {NoStop}%
\bibitem [{\citenamefont {Auld}(1959)}]{ferrite_2}%
  \BibitemOpen
  \bibfield  {author} {\bibinfo {author} {\bibfnamefont {B.}~\bibnamefont
  {Auld}},\ }\bibfield  {title} {\bibinfo {title} {The synthesis of symmetrical
  waveguide circulators},\ }\href {https://doi.org/10.1109/TMTT.1959.1124688}
  {\bibfield  {journal} {\bibinfo  {journal} {IRE Transactions on Microwave
  Theory and Techniques}\ }\textbf {\bibinfo {volume} {7}},\ \bibinfo {pages}
  {238} (\bibinfo {year} {1959})}\BibitemShut {NoStop}%
\bibitem [{\citenamefont {Wang}\ \emph {et~al.}(2021)\citenamefont {Wang},
  \citenamefont {van Geldern}, \citenamefont {Connolly}, \citenamefont {Wang},
  \citenamefont {Shilcusky}, \citenamefont {McDonald}, \citenamefont {Clerk},\
  and\ \citenamefont {Wang}}]{ferrite_3}%
  \BibitemOpen
  \bibfield  {author} {\bibinfo {author} {\bibfnamefont {Y.-Y.}\ \bibnamefont
  {Wang}}, \bibinfo {author} {\bibfnamefont {S.}~\bibnamefont {van Geldern}},
  \bibinfo {author} {\bibfnamefont {T.}~\bibnamefont {Connolly}}, \bibinfo
  {author} {\bibfnamefont {Y.-X.}\ \bibnamefont {Wang}}, \bibinfo {author}
  {\bibfnamefont {A.}~\bibnamefont {Shilcusky}}, \bibinfo {author}
  {\bibfnamefont {A.}~\bibnamefont {McDonald}}, \bibinfo {author}
  {\bibfnamefont {A.~A.}\ \bibnamefont {Clerk}},\ and\ \bibinfo {author}
  {\bibfnamefont {C.}~\bibnamefont {Wang}},\ }\bibfield  {title} {\bibinfo
  {title} {Low-loss ferrite circulator as a tunable chiral quantum system},\
  }\href {https://doi.org/10.1103/PhysRevApplied.16.064066} {\bibfield
  {journal} {\bibinfo  {journal} {Phys. Rev. Appl.}\ }\textbf {\bibinfo
  {volume} {16}},\ \bibinfo {pages} {064066} (\bibinfo {year}
  {2021})}\BibitemShut {NoStop}%
\bibitem [{\citenamefont {Hogan}(1952)}]{faraday_effect_1}%
  \BibitemOpen
  \bibfield  {author} {\bibinfo {author} {\bibfnamefont {C.~L.}\ \bibnamefont
  {Hogan}},\ }\bibfield  {title} {\bibinfo {title} {The ferromagnetic faraday
  effect at microwave frequencies and its applications: The microwave
  gyrator},\ }\href {https://doi.org/10.1002/j.1538-7305.1952.tb01374.x}
  {\bibfield  {journal} {\bibinfo  {journal} {The Bell System Technical
  Journal}\ }\textbf {\bibinfo {volume} {31}},\ \bibinfo {pages} {1} (\bibinfo
  {year} {1952})}\BibitemShut {NoStop}%
\bibitem [{\citenamefont {Aplet}\ and\ \citenamefont
  {Carson}(1964)}]{faraday_effect_2}%
  \BibitemOpen
  \bibfield  {author} {\bibinfo {author} {\bibfnamefont {L.~J.}\ \bibnamefont
  {Aplet}}\ and\ \bibinfo {author} {\bibfnamefont {J.~W.}\ \bibnamefont
  {Carson}},\ }\bibfield  {title} {\bibinfo {title} {A faraday effect optical
  isolator},\ }\href {https://doi.org/10.1364/AO.3.000544} {\bibfield
  {journal} {\bibinfo  {journal} {Appl. Opt.}\ }\textbf {\bibinfo {volume}
  {3}},\ \bibinfo {pages} {544} (\bibinfo {year} {1964})}\BibitemShut {NoStop}%
\bibitem [{\citenamefont {Gu}\ \emph {et~al.}(2017)\citenamefont {Gu},
  \citenamefont {Kockum}, \citenamefont {Miranowicz}, \citenamefont {xi~Liu},\
  and\ \citenamefont {Nori}}]{SQC_1}%
  \BibitemOpen
  \bibfield  {author} {\bibinfo {author} {\bibfnamefont {X.}~\bibnamefont
  {Gu}}, \bibinfo {author} {\bibfnamefont {A.~F.}\ \bibnamefont {Kockum}},
  \bibinfo {author} {\bibfnamefont {A.}~\bibnamefont {Miranowicz}}, \bibinfo
  {author} {\bibfnamefont {Y.}~\bibnamefont {xi~Liu}},\ and\ \bibinfo {author}
  {\bibfnamefont {F.}~\bibnamefont {Nori}},\ }\bibfield  {title} {\bibinfo
  {title} {Microwave photonics with superconducting quantum circuits},\ }\href
  {https://doi.org/https://doi.org/10.1016/j.physrep.2017.10.002} {\bibfield
  {journal} {\bibinfo  {journal} {Phys. Reports}\ }\textbf {\bibinfo {volume}
  {718-719}},\ \bibinfo {pages} {1} (\bibinfo {year} {2017})},\ \bibinfo {note}
  {microwave photonics with superconducting quantum circuits}\BibitemShut
  {NoStop}%
\bibitem [{\citenamefont {Rasmussen}\ \emph {et~al.}(2021)\citenamefont
  {Rasmussen}, \citenamefont {Christensen}, \citenamefont {Pedersen},
  \citenamefont {Kristensen}, \citenamefont {B\ae{}kkegaard}, \citenamefont
  {Loft},\ and\ \citenamefont {Zinner}}]{SQC_2}%
  \BibitemOpen
  \bibfield  {author} {\bibinfo {author} {\bibfnamefont {S.}~\bibnamefont
  {Rasmussen}}, \bibinfo {author} {\bibfnamefont {K.}~\bibnamefont
  {Christensen}}, \bibinfo {author} {\bibfnamefont {S.}~\bibnamefont
  {Pedersen}}, \bibinfo {author} {\bibfnamefont {L.}~\bibnamefont
  {Kristensen}}, \bibinfo {author} {\bibfnamefont {T.}~\bibnamefont
  {B\ae{}kkegaard}}, \bibinfo {author} {\bibfnamefont {N.}~\bibnamefont
  {Loft}},\ and\ \bibinfo {author} {\bibfnamefont {N.}~\bibnamefont {Zinner}},\
  }\bibfield  {title} {\bibinfo {title} {Superconducting circuit companion---an
  introduction with worked examples},\ }\href
  {https://doi.org/10.1103/PRXQuantum.2.040204} {\bibfield  {journal} {\bibinfo
   {journal} {PRX Quantum}\ }\textbf {\bibinfo {volume} {2}},\ \bibinfo {pages}
  {040204} (\bibinfo {year} {2021})}\BibitemShut {NoStop}%
\bibitem [{\citenamefont {Blais}\ \emph {et~al.}(2020)\citenamefont {Blais},
  \citenamefont {Girvin},\ and\ \citenamefont {Oliver}}]{SQC_3}%
  \BibitemOpen
  \bibfield  {author} {\bibinfo {author} {\bibfnamefont {A.}~\bibnamefont
  {Blais}}, \bibinfo {author} {\bibfnamefont {S.~M.}\ \bibnamefont {Girvin}},\
  and\ \bibinfo {author} {\bibfnamefont {W.~D.}\ \bibnamefont {Oliver}},\
  }\bibfield  {title} {\bibinfo {title} {Quantum information processing and
  quantum optics with circuit quantum electrodynamics},\ }\href
  {https://doi.org/10.1038/s41567-020-0806-z} {\bibfield  {journal} {\bibinfo
  {journal} {Nat. Physics}\ }\textbf {\bibinfo {volume} {16}},\ \bibinfo
  {pages} {247} (\bibinfo {year} {2020})}\BibitemShut {NoStop}%
\bibitem [{\citenamefont {Haroche}\ \emph {et~al.}(2020)\citenamefont
  {Haroche}, \citenamefont {Brune},\ and\ \citenamefont {Raimond}}]{SQC_4}%
  \BibitemOpen
  \bibfield  {author} {\bibinfo {author} {\bibfnamefont {S.}~\bibnamefont
  {Haroche}}, \bibinfo {author} {\bibfnamefont {M.}~\bibnamefont {Brune}},\
  and\ \bibinfo {author} {\bibfnamefont {J.~M.}\ \bibnamefont {Raimond}},\
  }\bibfield  {title} {\bibinfo {title} {From cavity to circuit quantum
  electrodynamics},\ }\href {https://doi.org/10.1038/s41567-020-0812-1}
  {\bibfield  {journal} {\bibinfo  {journal} {Nat. Physics}\ }\textbf {\bibinfo
  {volume} {16}},\ \bibinfo {pages} {243} (\bibinfo {year} {2020})}\BibitemShut
  {NoStop}%
\bibitem [{\citenamefont {Blais}\ \emph {et~al.}(2004)\citenamefont {Blais},
  \citenamefont {Huang}, \citenamefont {Wallraff}, \citenamefont {Girvin},\
  and\ \citenamefont {Schoelkopf}}]{SQC_5}%
  \BibitemOpen
  \bibfield  {author} {\bibinfo {author} {\bibfnamefont {A.}~\bibnamefont
  {Blais}}, \bibinfo {author} {\bibfnamefont {R.-S.}\ \bibnamefont {Huang}},
  \bibinfo {author} {\bibfnamefont {A.}~\bibnamefont {Wallraff}}, \bibinfo
  {author} {\bibfnamefont {S.~M.}\ \bibnamefont {Girvin}},\ and\ \bibinfo
  {author} {\bibfnamefont {R.~J.}\ \bibnamefont {Schoelkopf}},\ }\bibfield
  {title} {\bibinfo {title} {Cavity quantum electrodynamics for superconducting
  electrical circuits: An architecture for quantum computation},\ }\href
  {https://doi.org/10.1103/PhysRevA.69.062320} {\bibfield  {journal} {\bibinfo
  {journal} {Phys. Rev. A}\ }\textbf {\bibinfo {volume} {69}},\ \bibinfo
  {pages} {062320} (\bibinfo {year} {2004})}\BibitemShut {NoStop}%
\bibitem [{\citenamefont {Eddins}\ \emph {et~al.}(2019)\citenamefont {Eddins},
  \citenamefont {Kreikebaum}, \citenamefont {Toyli}, \citenamefont
  {Levenson-Falk}, \citenamefont {Dove}, \citenamefont {Livingston},
  \citenamefont {Levitan}, \citenamefont {Govia}, \citenamefont {Clerk},\ and\
  \citenamefont {Siddiqi}}]{A_A_1}%
  \BibitemOpen
  \bibfield  {author} {\bibinfo {author} {\bibfnamefont {A.}~\bibnamefont
  {Eddins}}, \bibinfo {author} {\bibfnamefont {J.~M.}\ \bibnamefont
  {Kreikebaum}}, \bibinfo {author} {\bibfnamefont {D.~M.}\ \bibnamefont
  {Toyli}}, \bibinfo {author} {\bibfnamefont {E.~M.}\ \bibnamefont
  {Levenson-Falk}}, \bibinfo {author} {\bibfnamefont {A.}~\bibnamefont {Dove}},
  \bibinfo {author} {\bibfnamefont {W.~P.}\ \bibnamefont {Livingston}},
  \bibinfo {author} {\bibfnamefont {B.~A.}\ \bibnamefont {Levitan}}, \bibinfo
  {author} {\bibfnamefont {L.~C.~G.}\ \bibnamefont {Govia}}, \bibinfo {author}
  {\bibfnamefont {A.~A.}\ \bibnamefont {Clerk}},\ and\ \bibinfo {author}
  {\bibfnamefont {I.}~\bibnamefont {Siddiqi}},\ }\bibfield  {title} {\bibinfo
  {title} {High-efficiency measurement of an artificial atom embedded in a
  parametric amplifier},\ }\href {https://doi.org/10.1103/PhysRevX.9.011004}
  {\bibfield  {journal} {\bibinfo  {journal} {Phys. Rev. X}\ }\textbf {\bibinfo
  {volume} {9}},\ \bibinfo {pages} {011004} (\bibinfo {year}
  {2019})}\BibitemShut {NoStop}%
\bibitem [{\citenamefont {Hoi}\ \emph {et~al.}(2011)\citenamefont {Hoi},
  \citenamefont {Wilson}, \citenamefont {Johansson}, \citenamefont {Palomaki},
  \citenamefont {Peropadre},\ and\ \citenamefont {Delsing}}]{A_A_2}%
  \BibitemOpen
  \bibfield  {author} {\bibinfo {author} {\bibfnamefont {I.-C.}\ \bibnamefont
  {Hoi}}, \bibinfo {author} {\bibfnamefont {C.~M.}\ \bibnamefont {Wilson}},
  \bibinfo {author} {\bibfnamefont {G.}~\bibnamefont {Johansson}}, \bibinfo
  {author} {\bibfnamefont {T.}~\bibnamefont {Palomaki}}, \bibinfo {author}
  {\bibfnamefont {B.}~\bibnamefont {Peropadre}},\ and\ \bibinfo {author}
  {\bibfnamefont {P.}~\bibnamefont {Delsing}},\ }\bibfield  {title} {\bibinfo
  {title} {Demonstration of a single-photon router in the microwave regime},\
  }\href {https://doi.org/10.1103/PhysRevLett.107.073601} {\bibfield  {journal}
  {\bibinfo  {journal} {Phys. Rev. Lett.}\ }\textbf {\bibinfo {volume} {107}},\
  \bibinfo {pages} {073601} (\bibinfo {year} {2011})}\BibitemShut {NoStop}%
\bibitem [{\citenamefont {Bosman}\ \emph {et~al.}(2017)\citenamefont {Bosman},
  \citenamefont {Gely}, \citenamefont {Singh}, \citenamefont {Bruno},
  \citenamefont {Bothner},\ and\ \citenamefont {Steele}}]{A_A_3}%
  \BibitemOpen
  \bibfield  {author} {\bibinfo {author} {\bibfnamefont {S.~J.}\ \bibnamefont
  {Bosman}}, \bibinfo {author} {\bibfnamefont {M.~F.}\ \bibnamefont {Gely}},
  \bibinfo {author} {\bibfnamefont {V.}~\bibnamefont {Singh}}, \bibinfo
  {author} {\bibfnamefont {A.}~\bibnamefont {Bruno}}, \bibinfo {author}
  {\bibfnamefont {D.}~\bibnamefont {Bothner}},\ and\ \bibinfo {author}
  {\bibfnamefont {G.~A.}\ \bibnamefont {Steele}},\ }\bibfield  {title}
  {\bibinfo {title} {Multi-mode ultra-strong coupling in circuit quantum
  electrodynamics},\ }\href {https://doi.org/10.1038/s41534-017-0046-y}
  {\bibfield  {journal} {\bibinfo  {journal} {npj Quantum Information}\
  }\textbf {\bibinfo {volume} {3}},\ \bibinfo {pages} {46} (\bibinfo {year}
  {2017})}\BibitemShut {NoStop}%
\bibitem [{\citenamefont {Zhu}\ and\ \citenamefont {Jia}(2019)}]{LCC_1}%
  \BibitemOpen
  \bibfield  {author} {\bibinfo {author} {\bibfnamefont {Y.~T.}\ \bibnamefont
  {Zhu}}\ and\ \bibinfo {author} {\bibfnamefont {W.~Z.}\ \bibnamefont {Jia}},\
  }\bibfield  {title} {\bibinfo {title} {Single-photon quantum router in the
  microwave regime utilizing double superconducting resonators with tunable
  coupling},\ }\href {https://doi.org/10.1103/PhysRevA.99.063815} {\bibfield
  {journal} {\bibinfo  {journal} {Phys. Rev. A}\ }\textbf {\bibinfo {volume}
  {99}},\ \bibinfo {pages} {063815} (\bibinfo {year} {2019})}\BibitemShut
  {NoStop}%
\bibitem [{\citenamefont {Wang}\ \emph {et~al.}(2024)\citenamefont {Wang},
  \citenamefont {Zhao}, \citenamefont {Sun}, \citenamefont {Xu}, \citenamefont
  {Li}, \citenamefont {Zheng}, \citenamefont {Liu},\ and\ \citenamefont
  {Li}}]{LCC_2}%
  \BibitemOpen
  \bibfield  {author} {\bibinfo {author} {\bibfnamefont {H.}~\bibnamefont
  {Wang}}, \bibinfo {author} {\bibfnamefont {Y.-J.}\ \bibnamefont {Zhao}},
  \bibinfo {author} {\bibfnamefont {H.-C.}\ \bibnamefont {Sun}}, \bibinfo
  {author} {\bibfnamefont {X.-W.}\ \bibnamefont {Xu}}, \bibinfo {author}
  {\bibfnamefont {Y.}~\bibnamefont {Li}}, \bibinfo {author} {\bibfnamefont
  {Y.}~\bibnamefont {Zheng}}, \bibinfo {author} {\bibfnamefont
  {Q.}~\bibnamefont {Liu}},\ and\ \bibinfo {author} {\bibfnamefont
  {R.}~\bibnamefont {Li}},\ }\bibfield  {title} {\bibinfo {title} {Controlling
  the qubit-qubit coupling in the superconducting circuit with double-resonator
  couplers},\ }\href {https://doi.org/10.1103/PhysRevA.109.012601} {\bibfield
  {journal} {\bibinfo  {journal} {Phys. Rev. A}\ }\textbf {\bibinfo {volume}
  {109}},\ \bibinfo {pages} {012601} (\bibinfo {year} {2024})}\BibitemShut
  {NoStop}%
\bibitem [{\citenamefont {Ren}\ \emph {et~al.}(2023)\citenamefont {Ren},
  \citenamefont {Ma}, \citenamefont {Xie},\ and\ \citenamefont {Li}}]{LCC_3}%
  \BibitemOpen
  \bibfield  {author} {\bibinfo {author} {\bibfnamefont {Y.-L.}\ \bibnamefont
  {Ren}}, \bibinfo {author} {\bibfnamefont {S.-L.}\ \bibnamefont {Ma}},
  \bibinfo {author} {\bibfnamefont {J.-K.}\ \bibnamefont {Xie}},\ and\ \bibinfo
  {author} {\bibfnamefont {F.-L.}\ \bibnamefont {Li}},\ }\bibfield  {title}
  {\bibinfo {title} {Nonreciprocal photonic quantum router via synthetic
  magnetism},\ }\href {https://doi.org/10.1063/5.0152354} {\bibfield  {journal}
  {\bibinfo  {journal} {Applied Physics Letters}\ }\textbf {\bibinfo {volume}
  {122}},\ \bibinfo {pages} {244002} (\bibinfo {year} {2023})},\ \Eprint
  {https://arxiv.org/abs/https://pubs.aip.org/aip/apl/article-pdf/doi/10.1063/5.0152354/17997433/244002\_1\_5.0152354.pdf}
  {https://pubs.aip.org/aip/apl/article-pdf/doi/10.1063/5.0152354/17997433/244002\_1\_5.0152354.pdf}
  \BibitemShut {NoStop}%
\bibitem [{\citenamefont {Cai}\ \emph {et~al.}(2024)\citenamefont {Cai},
  \citenamefont {Ma}, \citenamefont {Liu}, \citenamefont {Guo}, \citenamefont
  {Tan},\ and\ \citenamefont {Liu}}]{LCC_4}%
  \BibitemOpen
  \bibfield  {author} {\bibinfo {author} {\bibfnamefont {Y.}~\bibnamefont
  {Cai}}, \bibinfo {author} {\bibfnamefont {K.-J.}\ \bibnamefont {Ma}},
  \bibinfo {author} {\bibfnamefont {J.}~\bibnamefont {Liu}}, \bibinfo {author}
  {\bibfnamefont {G.-F.}\ \bibnamefont {Guo}}, \bibinfo {author} {\bibfnamefont
  {L.}~\bibnamefont {Tan}},\ and\ \bibinfo {author} {\bibfnamefont {W.-M.}\
  \bibnamefont {Liu}},\ }\bibfield  {title} {\bibinfo {title} {Highly scalable
  quantum router with frequency-independent scattering spectra},\ }\href
  {https://doi.org/10.1088/1367-2630/ad8d74} {\bibfield  {journal} {\bibinfo
  {journal} {New Journal of Physics}\ }\textbf {\bibinfo {volume} {26}},\
  \bibinfo {pages} {113003} (\bibinfo {year} {2024})}\BibitemShut {NoStop}%
\bibitem [{\citenamefont {Peropadre}\ \emph {et~al.}(2013)\citenamefont
  {Peropadre}, \citenamefont {Zueco}, \citenamefont {Wulschner}, \citenamefont
  {Deppe}, \citenamefont {Marx}, \citenamefont {Gross},\ and\ \citenamefont
  {Garc\'{\i}a-Ripoll}}]{SQUID_1}%
  \BibitemOpen
  \bibfield  {author} {\bibinfo {author} {\bibfnamefont {B.}~\bibnamefont
  {Peropadre}}, \bibinfo {author} {\bibfnamefont {D.}~\bibnamefont {Zueco}},
  \bibinfo {author} {\bibfnamefont {F.}~\bibnamefont {Wulschner}}, \bibinfo
  {author} {\bibfnamefont {F.}~\bibnamefont {Deppe}}, \bibinfo {author}
  {\bibfnamefont {A.}~\bibnamefont {Marx}}, \bibinfo {author} {\bibfnamefont
  {R.}~\bibnamefont {Gross}},\ and\ \bibinfo {author} {\bibfnamefont {J.~J.}\
  \bibnamefont {Garc\'{\i}a-Ripoll}},\ }\bibfield  {title} {\bibinfo {title}
  {Tunable coupling engineering between superconducting resonators: From
  sidebands to effective gauge fields},\ }\href
  {https://doi.org/10.1103/PhysRevB.87.134504} {\bibfield  {journal} {\bibinfo
  {journal} {Phys. Rev. B}\ }\textbf {\bibinfo {volume} {87}},\ \bibinfo
  {pages} {134504} (\bibinfo {year} {2013})}\BibitemShut {NoStop}%
\bibitem [{\citenamefont {Allman}\ \emph {et~al.}(2014)\citenamefont {Allman},
  \citenamefont {Whittaker}, \citenamefont {Castellanos-Beltran}, \citenamefont
  {Cicak}, \citenamefont {da~Silva}, \citenamefont {DeFeo}, \citenamefont
  {Lecocq}, \citenamefont {Sirois}, \citenamefont {Teufel}, \citenamefont
  {Aumentado},\ and\ \citenamefont {Simmonds}}]{SQUID_2}%
  \BibitemOpen
  \bibfield  {author} {\bibinfo {author} {\bibfnamefont {M.~S.}\ \bibnamefont
  {Allman}}, \bibinfo {author} {\bibfnamefont {J.~D.}\ \bibnamefont
  {Whittaker}}, \bibinfo {author} {\bibfnamefont {M.}~\bibnamefont
  {Castellanos-Beltran}}, \bibinfo {author} {\bibfnamefont {K.}~\bibnamefont
  {Cicak}}, \bibinfo {author} {\bibfnamefont {F.}~\bibnamefont {da~Silva}},
  \bibinfo {author} {\bibfnamefont {M.~P.}\ \bibnamefont {DeFeo}}, \bibinfo
  {author} {\bibfnamefont {F.}~\bibnamefont {Lecocq}}, \bibinfo {author}
  {\bibfnamefont {A.}~\bibnamefont {Sirois}}, \bibinfo {author} {\bibfnamefont
  {J.~D.}\ \bibnamefont {Teufel}}, \bibinfo {author} {\bibfnamefont
  {J.}~\bibnamefont {Aumentado}},\ and\ \bibinfo {author} {\bibfnamefont
  {R.~W.}\ \bibnamefont {Simmonds}},\ }\bibfield  {title} {\bibinfo {title}
  {Tunable resonant and nonresonant interactions between a phase qubit and $lc$
  resonator},\ }\href {https://doi.org/10.1103/PhysRevLett.112.123601}
  {\bibfield  {journal} {\bibinfo  {journal} {Phys. Rev. Lett.}\ }\textbf
  {\bibinfo {volume} {112}},\ \bibinfo {pages} {123601} (\bibinfo {year}
  {2014})}\BibitemShut {NoStop}%
\bibitem [{\citenamefont {Chen}\ \emph {et~al.}(2014)\citenamefont {Chen},
  \citenamefont {Neill}, \citenamefont {Roushan}, \citenamefont {Leung},
  \citenamefont {Fang}, \citenamefont {Barends}, \citenamefont {Kelly},
  \citenamefont {Campbell}, \citenamefont {Chen}, \citenamefont {Chiaro},
  \citenamefont {Dunsworth}, \citenamefont {Jeffrey}, \citenamefont {Megrant},
  \citenamefont {Mutus}, \citenamefont {O'Malley}, \citenamefont {Quintana},
  \citenamefont {Sank}, \citenamefont {Vainsencher}, \citenamefont {Wenner},
  \citenamefont {White}, \citenamefont {Geller}, \citenamefont {Cleland},\ and\
  \citenamefont {Martinis}}]{SQUID_3}%
  \BibitemOpen
  \bibfield  {author} {\bibinfo {author} {\bibfnamefont {Y.}~\bibnamefont
  {Chen}}, \bibinfo {author} {\bibfnamefont {C.}~\bibnamefont {Neill}},
  \bibinfo {author} {\bibfnamefont {P.}~\bibnamefont {Roushan}}, \bibinfo
  {author} {\bibfnamefont {N.}~\bibnamefont {Leung}}, \bibinfo {author}
  {\bibfnamefont {M.}~\bibnamefont {Fang}}, \bibinfo {author} {\bibfnamefont
  {R.}~\bibnamefont {Barends}}, \bibinfo {author} {\bibfnamefont
  {J.}~\bibnamefont {Kelly}}, \bibinfo {author} {\bibfnamefont
  {B.}~\bibnamefont {Campbell}}, \bibinfo {author} {\bibfnamefont
  {Z.}~\bibnamefont {Chen}}, \bibinfo {author} {\bibfnamefont {B.}~\bibnamefont
  {Chiaro}}, \bibinfo {author} {\bibfnamefont {A.}~\bibnamefont {Dunsworth}},
  \bibinfo {author} {\bibfnamefont {E.}~\bibnamefont {Jeffrey}}, \bibinfo
  {author} {\bibfnamefont {A.}~\bibnamefont {Megrant}}, \bibinfo {author}
  {\bibfnamefont {J.~Y.}\ \bibnamefont {Mutus}}, \bibinfo {author}
  {\bibfnamefont {P.~J.~J.}\ \bibnamefont {O'Malley}}, \bibinfo {author}
  {\bibfnamefont {C.~M.}\ \bibnamefont {Quintana}}, \bibinfo {author}
  {\bibfnamefont {D.}~\bibnamefont {Sank}}, \bibinfo {author} {\bibfnamefont
  {A.}~\bibnamefont {Vainsencher}}, \bibinfo {author} {\bibfnamefont
  {J.}~\bibnamefont {Wenner}}, \bibinfo {author} {\bibfnamefont {T.~C.}\
  \bibnamefont {White}}, \bibinfo {author} {\bibfnamefont {M.~R.}\ \bibnamefont
  {Geller}}, \bibinfo {author} {\bibfnamefont {A.~N.}\ \bibnamefont
  {Cleland}},\ and\ \bibinfo {author} {\bibfnamefont {J.~M.}\ \bibnamefont
  {Martinis}},\ }\bibfield  {title} {\bibinfo {title} {Qubit architecture with
  high coherence and fast tunable coupling},\ }\href
  {https://doi.org/10.1103/PhysRevLett.113.220502} {\bibfield  {journal}
  {\bibinfo  {journal} {Phys. Rev. Lett.}\ }\textbf {\bibinfo {volume} {113}},\
  \bibinfo {pages} {220502} (\bibinfo {year} {2014})}\BibitemShut {NoStop}%
\bibitem [{\citenamefont {Granata}\ and\ \citenamefont
  {Vettoliere}(2016)}]{SQUID_4}%
  \BibitemOpen
  \bibfield  {author} {\bibinfo {author} {\bibfnamefont {C.}~\bibnamefont
  {Granata}}\ and\ \bibinfo {author} {\bibfnamefont {A.}~\bibnamefont
  {Vettoliere}},\ }\bibfield  {title} {\bibinfo {title} {Nano superconducting
  quantum interference device: A powerful tool for nanoscale investigations},\
  }\href {https://doi.org/https://doi.org/10.1016/j.physrep.2015.12.001}
  {\bibfield  {journal} {\bibinfo  {journal} {Physics Reports}\ }\textbf
  {\bibinfo {volume} {614}},\ \bibinfo {pages} {1} (\bibinfo {year} {2016})},\
  \bibinfo {note} {nano Superconducting Quantum Interference device: a powerful
  tool for nanoscale investigations}\BibitemShut {NoStop}%
\bibitem [{\citenamefont {Noh}\ \emph {et~al.}(2021)\citenamefont {Noh},
  \citenamefont {Kindseth},\ and\ \citenamefont {Chandrasekhar}}]{SQUID_5}%
  \BibitemOpen
  \bibfield  {author} {\bibinfo {author} {\bibfnamefont {T.}~\bibnamefont
  {Noh}}, \bibinfo {author} {\bibfnamefont {A.}~\bibnamefont {Kindseth}},\ and\
  \bibinfo {author} {\bibfnamefont {V.}~\bibnamefont {Chandrasekhar}},\
  }\bibfield  {title} {\bibinfo {title} {Nonlocal superconducting quantum
  interference device},\ }\href {https://doi.org/10.1103/PhysRevB.104.064503}
  {\bibfield  {journal} {\bibinfo  {journal} {Phys. Rev. B}\ }\textbf {\bibinfo
  {volume} {104}},\ \bibinfo {pages} {064503} (\bibinfo {year}
  {2021})}\BibitemShut {NoStop}%
\bibitem [{\citenamefont {Lu}\ \emph {et~al.}(2017)\citenamefont {Lu},
  \citenamefont {Chakram}, \citenamefont {Leung}, \citenamefont {Earnest},
  \citenamefont {Naik}, \citenamefont {Huang}, \citenamefont {Groszkowski},
  \citenamefont {Kapit}, \citenamefont {Koch},\ and\ \citenamefont
  {Schuster}}]{SQUID_6}%
  \BibitemOpen
  \bibfield  {author} {\bibinfo {author} {\bibfnamefont {Y.}~\bibnamefont
  {Lu}}, \bibinfo {author} {\bibfnamefont {S.}~\bibnamefont {Chakram}},
  \bibinfo {author} {\bibfnamefont {N.}~\bibnamefont {Leung}}, \bibinfo
  {author} {\bibfnamefont {N.}~\bibnamefont {Earnest}}, \bibinfo {author}
  {\bibfnamefont {R.~K.}\ \bibnamefont {Naik}}, \bibinfo {author}
  {\bibfnamefont {Z.}~\bibnamefont {Huang}}, \bibinfo {author} {\bibfnamefont
  {P.}~\bibnamefont {Groszkowski}}, \bibinfo {author} {\bibfnamefont
  {E.}~\bibnamefont {Kapit}}, \bibinfo {author} {\bibfnamefont
  {J.}~\bibnamefont {Koch}},\ and\ \bibinfo {author} {\bibfnamefont {D.~I.}\
  \bibnamefont {Schuster}},\ }\bibfield  {title} {\bibinfo {title} {Universal
  stabilization of a parametrically coupled qubit},\ }\href
  {https://doi.org/10.1103/PhysRevLett.119.150502} {\bibfield  {journal}
  {\bibinfo  {journal} {Phys. Rev. Lett.}\ }\textbf {\bibinfo {volume} {119}},\
  \bibinfo {pages} {150502} (\bibinfo {year} {2017})}\BibitemShut {NoStop}%
\bibitem [{\citenamefont {Ranzani}\ and\ \citenamefont
  {Aumentado}(2015)}]{nonreciprocal_graph}%
  \BibitemOpen
  \bibfield  {author} {\bibinfo {author} {\bibfnamefont {L.}~\bibnamefont
  {Ranzani}}\ and\ \bibinfo {author} {\bibfnamefont {J.}~\bibnamefont
  {Aumentado}},\ }\bibfield  {title} {\bibinfo {title} {Graph-based analysis of
  nonreciprocity in coupled-mode systems},\ }\href
  {https://doi.org/10.1088/1367-2630/17/2/023024} {\bibfield  {journal}
  {\bibinfo  {journal} {New Journal of Physics}\ }\textbf {\bibinfo {volume}
  {17}},\ \bibinfo {pages} {023024} (\bibinfo {year} {2015})}\BibitemShut
  {NoStop}%
\bibitem [{\citenamefont {Xu}\ \emph {et~al.}(2017)\citenamefont {Xu},
  \citenamefont {Chen}, \citenamefont {Li},\ and\ \citenamefont
  {Liu}}]{phase_included}%
  \BibitemOpen
  \bibfield  {author} {\bibinfo {author} {\bibfnamefont {X.-W.}\ \bibnamefont
  {Xu}}, \bibinfo {author} {\bibfnamefont {A.-X.}\ \bibnamefont {Chen}},
  \bibinfo {author} {\bibfnamefont {Y.}~\bibnamefont {Li}},\ and\ \bibinfo
  {author} {\bibfnamefont {Y.-x.}\ \bibnamefont {Liu}},\ }\bibfield  {title}
  {\bibinfo {title} {Single-photon nonreciprocal transport in one-dimensional
  coupled-resonator waveguides},\ }\href
  {https://doi.org/10.1103/PhysRevA.95.063808} {\bibfield  {journal} {\bibinfo
  {journal} {Phys. Rev. A}\ }\textbf {\bibinfo {volume} {95}},\ \bibinfo
  {pages} {063808} (\bibinfo {year} {2017})}\BibitemShut {NoStop}%
\bibitem [{\citenamefont {Vasko}(2017)}]{TL_1}%
  \BibitemOpen
  \bibfield  {author} {\bibinfo {author} {\bibfnamefont {F.~T.}\ \bibnamefont
  {Vasko}},\ }\bibfield  {title} {\bibinfo {title} {Flux noise in a
  superconducting transmission line},\ }\href
  {https://doi.org/10.1103/PhysRevApplied.8.024003} {\bibfield  {journal}
  {\bibinfo  {journal} {Phys. Rev. Appl.}\ }\textbf {\bibinfo {volume} {8}},\
  \bibinfo {pages} {024003} (\bibinfo {year} {2017})}\BibitemShut {NoStop}%
\bibitem [{\citenamefont {Ren}\ \emph {et~al.}(2022)\citenamefont {Ren},
  \citenamefont {Ma}, \citenamefont {Xie}, \citenamefont {Li}, \citenamefont
  {Cao},\ and\ \citenamefont {Li}}]{nonreciprocal_router}%
  \BibitemOpen
  \bibfield  {author} {\bibinfo {author} {\bibfnamefont {Y.-l.}\ \bibnamefont
  {Ren}}, \bibinfo {author} {\bibfnamefont {S.-l.}\ \bibnamefont {Ma}},
  \bibinfo {author} {\bibfnamefont {J.-k.}\ \bibnamefont {Xie}}, \bibinfo
  {author} {\bibfnamefont {X.-k.}\ \bibnamefont {Li}}, \bibinfo {author}
  {\bibfnamefont {M.-t.}\ \bibnamefont {Cao}},\ and\ \bibinfo {author}
  {\bibfnamefont {F.-l.}\ \bibnamefont {Li}},\ }\bibfield  {title} {\bibinfo
  {title} {Nonreciprocal single-photon quantum router},\ }\href
  {https://doi.org/10.1103/PhysRevA.105.013711} {\bibfield  {journal} {\bibinfo
   {journal} {Phys. Rev. A}\ }\textbf {\bibinfo {volume} {105}},\ \bibinfo
  {pages} {013711} (\bibinfo {year} {2022})}\BibitemShut {NoStop}%
\bibitem [{\citenamefont {Koch}\ \emph {et~al.}(2010)\citenamefont {Koch},
  \citenamefont {Houck}, \citenamefont {Hur},\ and\ \citenamefont
  {Girvin}}]{TRSB_1}%
  \BibitemOpen
  \bibfield  {author} {\bibinfo {author} {\bibfnamefont {J.}~\bibnamefont
  {Koch}}, \bibinfo {author} {\bibfnamefont {A.~A.}\ \bibnamefont {Houck}},
  \bibinfo {author} {\bibfnamefont {K.~L.}\ \bibnamefont {Hur}},\ and\ \bibinfo
  {author} {\bibfnamefont {S.~M.}\ \bibnamefont {Girvin}},\ }\bibfield  {title}
  {\bibinfo {title} {Time-reversal-symmetry breaking in circuit-qed-based
  photon lattices},\ }\href {https://doi.org/10.1103/PhysRevA.82.043811}
  {\bibfield  {journal} {\bibinfo  {journal} {Phys. Rev. A}\ }\textbf {\bibinfo
  {volume} {82}},\ \bibinfo {pages} {043811} (\bibinfo {year}
  {2010})}\BibitemShut {NoStop}%
\bibitem [{\citenamefont {Sliwa}\ \emph {et~al.}(2015)\citenamefont {Sliwa},
  \citenamefont {Hatridge}, \citenamefont {Narla}, \citenamefont {Shankar},
  \citenamefont {Frunzio}, \citenamefont {Schoelkopf},\ and\ \citenamefont
  {Devoret}}]{TRSB_2}%
  \BibitemOpen
  \bibfield  {author} {\bibinfo {author} {\bibfnamefont {K.~M.}\ \bibnamefont
  {Sliwa}}, \bibinfo {author} {\bibfnamefont {M.}~\bibnamefont {Hatridge}},
  \bibinfo {author} {\bibfnamefont {A.}~\bibnamefont {Narla}}, \bibinfo
  {author} {\bibfnamefont {S.}~\bibnamefont {Shankar}}, \bibinfo {author}
  {\bibfnamefont {L.}~\bibnamefont {Frunzio}}, \bibinfo {author} {\bibfnamefont
  {R.~J.}\ \bibnamefont {Schoelkopf}},\ and\ \bibinfo {author} {\bibfnamefont
  {M.~H.}\ \bibnamefont {Devoret}},\ }\bibfield  {title} {\bibinfo {title}
  {Reconfigurable josephson circulator/directional amplifier},\ }\href
  {https://doi.org/10.1103/PhysRevX.5.041020} {\bibfield  {journal} {\bibinfo
  {journal} {Phys. Rev. X}\ }\textbf {\bibinfo {volume} {5}},\ \bibinfo {pages}
  {041020} (\bibinfo {year} {2015})}\BibitemShut {NoStop}%
\bibitem [{\citenamefont {Megrant}\ \emph {et~al.}(2012)\citenamefont
  {Megrant}, \citenamefont {Neill}, \citenamefont {Barends}, \citenamefont
  {Chiaro}, \citenamefont {Chen}, \citenamefont {Feigl}, \citenamefont {Kelly},
  \citenamefont {Lucero}, \citenamefont {Mariantoni}, \citenamefont
  {O’Malley}, \citenamefont {Sank}, \citenamefont {Vainsencher},
  \citenamefont {Wenner}, \citenamefont {White}, \citenamefont {Yin},
  \citenamefont {Zhao}, \citenamefont {Palmstrøm}, \citenamefont {Martinis},\
  and\ \citenamefont {Cleland}}]{Hcavity_1}%
  \BibitemOpen
  \bibfield  {author} {\bibinfo {author} {\bibfnamefont {A.}~\bibnamefont
  {Megrant}}, \bibinfo {author} {\bibfnamefont {C.}~\bibnamefont {Neill}},
  \bibinfo {author} {\bibfnamefont {R.}~\bibnamefont {Barends}}, \bibinfo
  {author} {\bibfnamefont {B.}~\bibnamefont {Chiaro}}, \bibinfo {author}
  {\bibfnamefont {Y.}~\bibnamefont {Chen}}, \bibinfo {author} {\bibfnamefont
  {L.}~\bibnamefont {Feigl}}, \bibinfo {author} {\bibfnamefont
  {J.}~\bibnamefont {Kelly}}, \bibinfo {author} {\bibfnamefont
  {E.}~\bibnamefont {Lucero}}, \bibinfo {author} {\bibfnamefont
  {M.}~\bibnamefont {Mariantoni}}, \bibinfo {author} {\bibfnamefont {P.~J.~J.}\
  \bibnamefont {O’Malley}}, \bibinfo {author} {\bibfnamefont
  {D.}~\bibnamefont {Sank}}, \bibinfo {author} {\bibfnamefont {A.}~\bibnamefont
  {Vainsencher}}, \bibinfo {author} {\bibfnamefont {J.}~\bibnamefont {Wenner}},
  \bibinfo {author} {\bibfnamefont {T.~C.}\ \bibnamefont {White}}, \bibinfo
  {author} {\bibfnamefont {Y.}~\bibnamefont {Yin}}, \bibinfo {author}
  {\bibfnamefont {J.}~\bibnamefont {Zhao}}, \bibinfo {author} {\bibfnamefont
  {C.~J.}\ \bibnamefont {Palmstrøm}}, \bibinfo {author} {\bibfnamefont
  {J.~M.}\ \bibnamefont {Martinis}},\ and\ \bibinfo {author} {\bibfnamefont
  {A.~N.}\ \bibnamefont {Cleland}},\ }\bibfield  {title} {\bibinfo {title}
  {Planar superconducting resonators with internal quality factors above one
  million},\ }\href {https://doi.org/10.1063/1.3693409} {\bibfield  {journal}
  {\bibinfo  {journal} {Applied Physics Letters}\ }\textbf {\bibinfo {volume}
  {100}},\ \bibinfo {pages} {113510} (\bibinfo {year} {2012})}\BibitemShut
  {NoStop}%
\bibitem [{\citenamefont {Romanenko}\ \emph {et~al.}(2020)\citenamefont
  {Romanenko}, \citenamefont {Pilipenko}, \citenamefont {Zorzetti},
  \citenamefont {Frolov}, \citenamefont {Awida}, \citenamefont {Belomestnykh},
  \citenamefont {Posen},\ and\ \citenamefont {Grassellino}}]{Hcavity_2}%
  \BibitemOpen
  \bibfield  {author} {\bibinfo {author} {\bibfnamefont {A.}~\bibnamefont
  {Romanenko}}, \bibinfo {author} {\bibfnamefont {R.}~\bibnamefont
  {Pilipenko}}, \bibinfo {author} {\bibfnamefont {S.}~\bibnamefont {Zorzetti}},
  \bibinfo {author} {\bibfnamefont {D.}~\bibnamefont {Frolov}}, \bibinfo
  {author} {\bibfnamefont {M.}~\bibnamefont {Awida}}, \bibinfo {author}
  {\bibfnamefont {S.}~\bibnamefont {Belomestnykh}}, \bibinfo {author}
  {\bibfnamefont {S.}~\bibnamefont {Posen}},\ and\ \bibinfo {author}
  {\bibfnamefont {A.}~\bibnamefont {Grassellino}},\ }\bibfield  {title}
  {\bibinfo {title} {Three-dimensional superconducting resonators at t < 20mk
  with photon lifetimes up to t = 2s},\ }\href
  {https://doi.org/10.1103/PhysRevApplied.13.034032} {\bibfield  {journal}
  {\bibinfo  {journal} {Phys. Rev. Appl.}\ }\textbf {\bibinfo {volume} {13}},\
  \bibinfo {pages} {034032} (\bibinfo {year} {2020})}\BibitemShut {NoStop}%
\bibitem [{\citenamefont {J\"ahne}\ \emph {et~al.}(2009)\citenamefont
  {J\"ahne}, \citenamefont {Genes}, \citenamefont {Hammerer}, \citenamefont
  {Wallquist}, \citenamefont {Polzik},\ and\ \citenamefont
  {Zoller}}]{adiabatically_eliminating_1}%
  \BibitemOpen
  \bibfield  {author} {\bibinfo {author} {\bibfnamefont {K.}~\bibnamefont
  {J\"ahne}}, \bibinfo {author} {\bibfnamefont {C.}~\bibnamefont {Genes}},
  \bibinfo {author} {\bibfnamefont {K.}~\bibnamefont {Hammerer}}, \bibinfo
  {author} {\bibfnamefont {M.}~\bibnamefont {Wallquist}}, \bibinfo {author}
  {\bibfnamefont {E.~S.}\ \bibnamefont {Polzik}},\ and\ \bibinfo {author}
  {\bibfnamefont {P.}~\bibnamefont {Zoller}},\ }\bibfield  {title} {\bibinfo
  {title} {Cavity-assisted squeezing of a mechanical oscillator},\ }\href
  {https://doi.org/10.1103/PhysRevA.79.063819} {\bibfield  {journal} {\bibinfo
  {journal} {Phys. Rev. A}\ }\textbf {\bibinfo {volume} {79}},\ \bibinfo
  {pages} {063819} (\bibinfo {year} {2009})}\BibitemShut {NoStop}%
\bibitem [{\citenamefont {Xu}\ \emph {et~al.}(2015)\citenamefont {Xu},
  \citenamefont {Liu}, \citenamefont {Sun},\ and\ \citenamefont
  {Li}}]{adiabatically_eliminating_2}%
  \BibitemOpen
  \bibfield  {author} {\bibinfo {author} {\bibfnamefont {X.-W.}\ \bibnamefont
  {Xu}}, \bibinfo {author} {\bibfnamefont {Y.-x.}\ \bibnamefont {Liu}},
  \bibinfo {author} {\bibfnamefont {C.-P.}\ \bibnamefont {Sun}},\ and\ \bibinfo
  {author} {\bibfnamefont {Y.}~\bibnamefont {Li}},\ }\bibfield  {title}
  {\bibinfo {title} {Mechanical $\mathcal{PT}$ symmetry in coupled
  optomechanical systems},\ }\href {https://doi.org/10.1103/PhysRevA.92.013852}
  {\bibfield  {journal} {\bibinfo  {journal} {Phys. Rev. A}\ }\textbf {\bibinfo
  {volume} {92}},\ \bibinfo {pages} {013852} (\bibinfo {year}
  {2015})}\BibitemShut {NoStop}%
\bibitem [{\citenamefont {Xu}\ \emph {et~al.}(2016)\citenamefont {Xu},
  \citenamefont {Li}, \citenamefont {Chen},\ and\ \citenamefont
  {Liu}}]{adiabatically_eliminating_3}%
  \BibitemOpen
  \bibfield  {author} {\bibinfo {author} {\bibfnamefont {X.-W.}\ \bibnamefont
  {Xu}}, \bibinfo {author} {\bibfnamefont {Y.}~\bibnamefont {Li}}, \bibinfo
  {author} {\bibfnamefont {A.-X.}\ \bibnamefont {Chen}},\ and\ \bibinfo
  {author} {\bibfnamefont {Y.-x.}\ \bibnamefont {Liu}},\ }\bibfield  {title}
  {\bibinfo {title} {Nonreciprocal conversion between microwave and optical
  photons in electro-optomechanical systems},\ }\href
  {https://doi.org/10.1103/PhysRevA.93.023827} {\bibfield  {journal} {\bibinfo
  {journal} {Phys. Rev. A}\ }\textbf {\bibinfo {volume} {93}},\ \bibinfo
  {pages} {023827} (\bibinfo {year} {2016})}\BibitemShut {NoStop}%
\bibitem [{\citenamefont {Liu}\ \emph {et~al.}(2005)\citenamefont {Liu},
  \citenamefont {You}, \citenamefont {Wei}, \citenamefont {Sun},\ and\
  \citenamefont {Nori}}]{selection_rule}%
  \BibitemOpen
  \bibfield  {author} {\bibinfo {author} {\bibfnamefont {Y.-x.}\ \bibnamefont
  {Liu}}, \bibinfo {author} {\bibfnamefont {J.~Q.}\ \bibnamefont {You}},
  \bibinfo {author} {\bibfnamefont {L.~F.}\ \bibnamefont {Wei}}, \bibinfo
  {author} {\bibfnamefont {C.~P.}\ \bibnamefont {Sun}},\ and\ \bibinfo {author}
  {\bibfnamefont {F.}~\bibnamefont {Nori}},\ }\bibfield  {title} {\bibinfo
  {title} {Optical selection rules and phase-dependent adiabatic state control
  in a superconducting quantum circuit},\ }\href
  {https://doi.org/10.1103/PhysRevLett.95.087001} {\bibfield  {journal}
  {\bibinfo  {journal} {Phys. Rev. Lett.}\ }\textbf {\bibinfo {volume} {95}},\
  \bibinfo {pages} {087001} (\bibinfo {year} {2005})}\BibitemShut {NoStop}%
\end{thebibliography}%

\end{document}